\documentclass[12pt,preprint]{aastex} 		

\usepackage{natbib}
\usepackage{mathptm}   

\newcommand{\kms}{km s$^{-1}$}
\newcommand{\solarmass}{\ensuremath{ \mathnormal{M}_{\Sun} }}
\newcommand{\mass}{\ensuremath{ \mathnormal{M} }}

\begin{document}

\title{Why are the K dwarfs in the Pleiades so Blue?\altaffilmark{1,2}}

\slugcomment{Accepted to the Astronomical Journal} 
\shortauthors{Stauffer et al.} 
\shorttitle{K dwarfs in the Pleiades}

\author{John R. Stauffer}
\affil{SIRTF Science Center, Caltech 314-6, Pasadena, CA  91125}

\author{Burton F. Jones}
\affil{Lick Observatory, University of California, Santa Cruz, CA 95064}

\author{Dana Backman} 
\affil{Department of Physics and
Astronomy, Franklin \& Marshall College, PO Box 3003,
Lancaster, PA 17604}

\author{Lee W. Hartmann}
\affil{Smithsonian Astrophysical Observatory, 60 Garden St.,
Cambridge, MA 02138}

\author{David Barrado y Navascu{\'e}s}
\affil{Laboratorio de Astrof\'isica Espacial y F\'isica Fundamental,
Apdo. 50727, 28080 Madrid. Spain }

\author{Marc H. Pinsonneault and Donald M. Terndrup}
\affil{Department of Astronomy, The Ohio State University,
140 West 18th Avenue, Columbus, OH 43210}

\author{August A. Muench}
\affil{SIRTF Science Center, Caltech 314-6, Pasadena, CA  91125}

\altaffiltext{1}{Some of the data presented herein were obtained
at the W.M. Keck Observatory, which is operated as a scientific
partnership among the California Institute of Technology,
the University of California and the National Aeronautics and
Space Administration. The Observatory was made possible by
the generous financial support of the W.M. Keck Foundation.} 
\altaffiltext{2}{This publication makes use of data products 
from the Two Micron All Sky Survey, which is a joint project 
of the University of Massachusetts and the Infrared Processing 
and Analysis Center/California Institute of Technology, 
funded by the National Aeronautics and Space Administration 
and the National Science Foundation.}

\begin{abstract}

The K dwarfs in the Pleiades fall nearly one half
magnitude below a main sequence isochrone when plotted
in a color-magnitude diagram utilizing $V$ magnitude as
the luminosity index and $B-V$ as the color index.  This
peculiarity has been known for forty years but has gone
unexplained and mostly ignored.  When compared to Praesepe
members, the Pleiades K dwarfs again are subluminous (or
blue) in a color-magnitude diagram using $B-V$ as the color
index.  However, using $V-I$ as the color index, stars in
the two clusters are coincident to M$_V$ $\sim$ 10; using
$V-K$ as the color index, Pleiades late K and M stars
fall above the main sequence locus defined by Praesepe
members.  We believe that the anomalous spectral energy
distributions for the Pleiades K dwarfs, as compared to older
clusters, are a consequence of rapid stellar rotation and
may be primarily due to spottedness.  If so, the required
areal filling factor for the cool component has to be very large
($\geq$ 50\%).  Weak-lined T Tauri stars have similar
color anomalies, and we suspect this is a common feature of
all very young K dwarfs (sp. type $>$ K3).  The peculiar
spectral energy distribution needs to be considered in
deriving accurate pre-main sequence isochrone-fitting ages
for clusters like the Pleiades since the age derived will
depend on the temperature index used.

\end{abstract}


\section{Introduction}
\label{sec:intro}

Almost forty years ago, \citet{herbig62} used
photometry of low mass stars in the Pleiades obtained by
\citet[][hereafter JM]{jm58} as the primary observational
basis for his ``bi-modal star-formation'' model.  In that
model, low-mass stars form continuously in a molecular
cloud, whereas the high mass stars form later as a result
of some external event such as a nearby supernova or
a spiral-arm density wave passage.  The evidence for
this was that the age inferred for the Pleiades from
the location of the pre-main sequence turnon point was
considerably older than the age inferred for the B stars
from comparison to post-main sequence tracks.  In other
words, the Pleiades K and early-M dwarfs were fainter,
or bluer, than expected for the adopted upper-main
sequence turnoff age of the cluster.  Ten years later,
\citet{jones72} utilized the same observational data
from JM to show that, in fact, the Pleiades K dwarfs
fall well below the ZAMS in an M$_V$ vs.\ $B-V$
color-magnitude diagram (CMD).
Jones argued that this anomaly might be the signature of
neutral extinction dust shells. 
\citet{poveda65} had also argued for the presence of such
neutral extinction dust shells surrounding Pleiades low mass stars. 
However, no subsequent evidence to verify the presence of
such dust shells has been published. 
New photometry of the Pleiades by \citet{landolt79} and
\citet{stauffer80}, also indicated the apparent displacement 
of the Pleiades K dwarfs below the ZAMS when using $B-V$ 
as the temperature index; however, \citeauthor{stauffer80}
was also able to show that the $V$ band deficits did not correlate
with $K$-band IR-excesses, thus casting doubt on the dust shell
hypothesis (dependent on the precise shell geometry).

Approximately another decade later, \citet{mermilliod92} 
rediscovered the apparent under-luminosity of the Pleiades K dwarfs.  
While attempting to fit a ZAMS M$_V$ vs. $B-V$ curve to the lower 
main sequence of the Pleiades in order to identify those 
cluster members that are photometric binaries, those authors 
noted that, ``the fit to the colour-magnitude diagram was 
not very good, expecially for (B $-$ V) $>$ 0.70 where the 
ZAMS became brighter than the observed data at a given colour.  
In fact, the difference between the ZAMS and the data was as 
much as 0.5 magnitudes at (B $-$ V )$_o$\ = 1.3.''  Because
this discrepancy was not directly relevant to the goals of
that paper, they chose simply to redefine the ZAMS curve so 
that it traced the lower envelope of the Pleiades distribution.

In the 30 years since \citet{jones72}, no paper has explained 
the mystery pointed out there, and in general that anomaly has
largely been ignored.  It remains unclear whether the anomaly
is due to a flux deficit at yellow wavelengths or an excess
flux in the blue.  The apparent age
difference between high and low mass stars in the Pleiades
has continued to be cited as evidence for more refined
bimodal star-formation models \citep{ns80,larson82},
which remains to this day as a commonly held belief for
the time history of star formation in a molecular cloud.
Reddening estimates for the Pleiades K and M dwarfs are
usually calculated assuming those stars have normal $B-V$
colors for their spectral types \citep[e.g.,][]{breger86}
which could well be incorrect, possibly leading to an
underestimate of the reddening for these stars.  Similarly,
for metallicity and lithium abundance derivations,
effective temperatures are often calculated using $B-V$
as a temperature index, which could lead to effective
temperatures which are too warm and hence abundances which
are too high.

An important clue to the resolution of this problem was provided
by \citet[][ hereafter vLAM87]{vleeuwen87}. Those authors compared
Pleiades K dwarfs to field K dwarfs in two-color diagrams using
Walraven $VBLUW$ photometry, and showed that the Pleiades
K dwarfs are too blue in $U-B$ or $U-L$ for their $B-V$ color
(where $U$ and $L$ are Walraven filters centered at about
$3620\;\AA$\ and $3825\;\AA$, respectively).
Based on their photometry, they concluded that,
``The blue-excess of the Pleiades K dwarfs starts at the
Balmer jump, and is, considering the fact that many of these
stars have active chromospheres producing H$\alpha$\ and 
H$\beta$\ emission, probably due to Balmer-continuum radiation.
There is some indication ... that the fast rotating stars are
more active than the slow rotating ones, as would be expected
and as was also observed by \citet{shs84}.''  
vLAM87 did not make reference to the \citet{herbig62} or
\citet{jones72} papers, and did not mention the unusual location
of the Pleiades K dwarfs in the CMD compared to less active/older stars. 
However, if the observed ultraviolet excess extended into the $B$ band, 
it could provide an explanation for the quandary posed by \citet{jones72}.

We have several goals for the current paper.  First,
we will review all available optical photometry for the
low-mass Pleiades members in order to document the K dwarf
CMD anomaly and illustrate that it is not simply due to an
unexpected error in the original JM photometry.  Second,
we will consider other color-magnitude and color-color diagrams
for the Pleiades as a means to determine whether the \citet{jones72} effect
is a flux deficit at $V$ or an excess in $B$ producing
bluer $B-V$ colors (it is, in fact, the latter).  Third,
we will examine photometry for other young open clusters
in order to constrain any physical model for the effect.
We provide new spectra of a sample of the Pleiades K dwarfs
at blue wavelengths in order to define better the exact
wavelength dependence of the effect.  Finally, we explore
possible physical explanations which could in principle
produce the blue colors of the faint Pleiades stars.

\section{Archival Evidence for Unusual Colors of Pleiades K Dwarfs}
\label{sec:archival}

\begin{figure} %
    \includegraphics[angle=00,totalheight=6.5in]{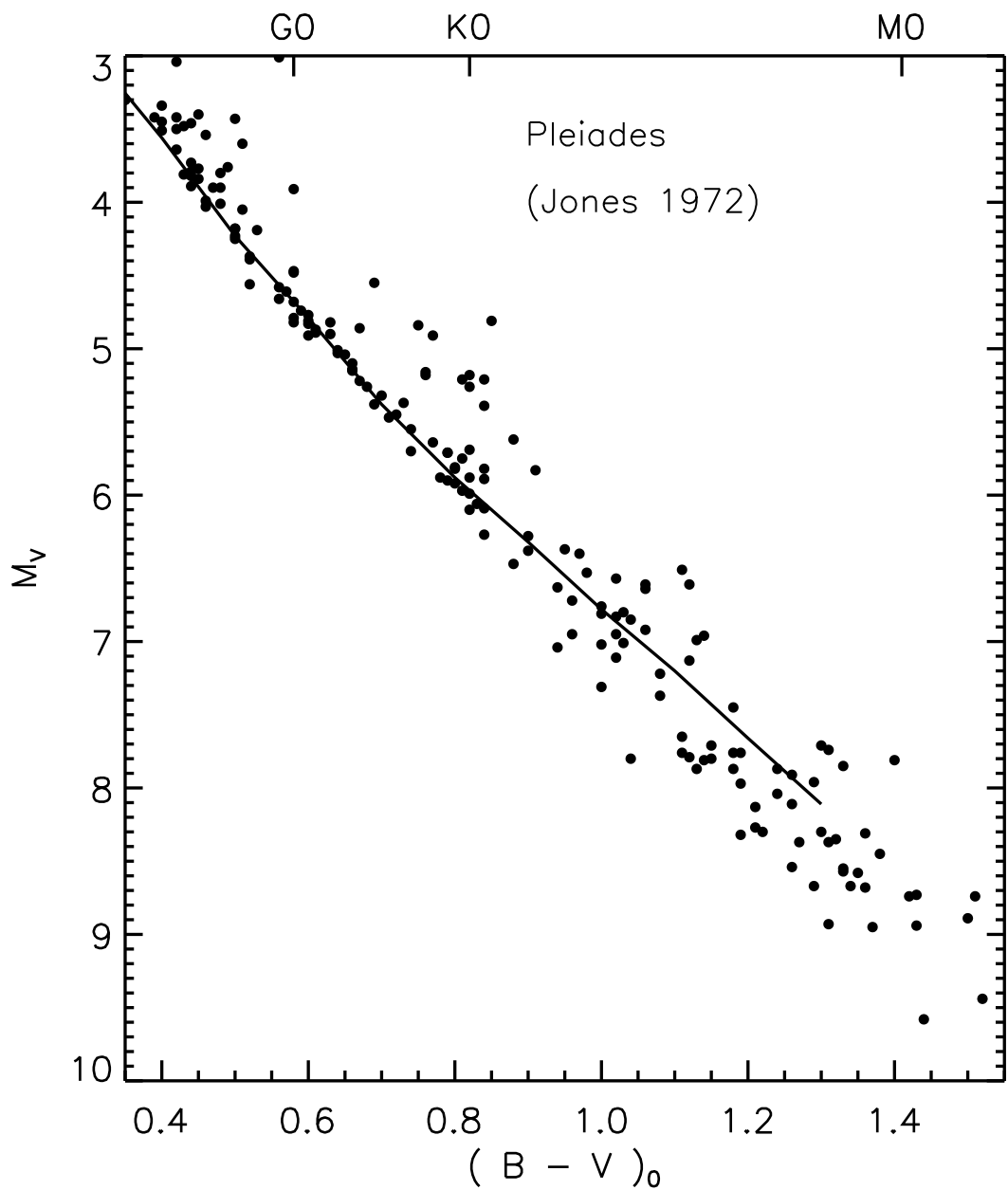}
    \caption{
	$V$ versus $B-V$ color-magnitude diagram for Pleiades members derived
        from photoelectric photometry by Johnson \& Mitchell and Jones \& Iriarte. 
	The figure is essentially the same as Figure 1 of Jones (1972), where
        in this case we assume a distance modulus of $5.53$ and $E(B-V)=0.04$.
	The curve is the main sequence locus adopted by Jones, from \citet{ji58}
	and we indicate relevant spectral types along the top axis for clarity.
    \label{fig:jonesbvv}
    }
\end{figure}

\citet{jones72} provided compelling evidence that the
Pleiades K dwarfs have abnormal photometric properties
based on the location of those stars in a CMD compared to
a zero-age main sequence derived by \citet{ji58}.
In addition to the JM photometry, Jones also included new
observations obtained by himself and by \citet{iriarte67}.
We provide a new version of Jones' color-magnitude
diagram here as our Figure \ref{fig:jonesbvv}.
As noted by Jones, the figure shows that the Pleiades K dwarfs
($B-V \gtrsim 1.0$) are displaced on average 0.5 mag below the
ZAMS, whereas the F and G stars essentially fall along the ZAMS.
Note that one expects $\sim$\ 20\% of the low mass Pleiades
members to be binaries with nearly equal mass components,
and those binaries would fall $\sim 0.7$ mag above the
single star locus, thus explaining much of the scatter
above the lower envelope of stars shown in
Figure \ref{fig:jonesbvv}.

Given that the photometry from Johnson and his
collaborators defined the $UBV$ system, it is unlikely a
priori that the above effect is due to a systematic error
in the JM Pleiades data.  Even so, we have attempted to examine
all other relevant photometry in order to verify the reality of
the effect illustrated in Figure \ref{fig:jonesbvv}. 
We provide the results of that literature survey in the
next several figures.

First, is the main sequence curve shown in Figure \ref{fig:jonesbvv}
an accurate rendition of the ZAMS?  Figure \ref{fig:hippbvv}
provides a plot of data from the Hipparcos catalog \citep{hipparcos}
for field stars within 25 pc, where the curve shown is the Johnson \&
Iriarte ZAMS used in Figure \ref{fig:jonesbvv}.  Because the
stars in Figure \ref{fig:hippbvv} include single and binary stars
of ages up to 10 Gyr and stars
with a  range of metallicity, the solar metallicity ZAMS should
roughly fall along a lower envelope to the density maximum in the
figure. Our judgment is that the Johnson \& Iriarte curve
traces that locus nearly as accurately as can be done,
particularly for $B-V > 1.0$.

Figure \ref{fig:otherbvv} shows independent photometry from several
papers that provided photometry for low mass stars in
the Pleiades \citep{landolt79, stauffer80, stauffer82a,
stauffer84, vleeuwen86, messina01}.  We have transformed
the V$_W$\ and (v-b)$_W$\ photometry obtained by \citet{vleeuwen86}
into V$_J$\ and (B-V)$_J$ using the formulae provided in
\citet{mermilliod92}. All of these data sets corroborate
the displacement of the K dwarfs beneath the ZAMS, and by a
similar degree. Therefore, we conclude that the anomaly is real. 

In fact, the previous figures underemphasize the true difference
between the Pleiades K dwarf locus and the nominal ZAMS for two
reasons: differential reddening within the Pleiades and binarity.  
Subsequent to the \citet{jones72} paper, it was realized that there is
a small molecular cloud currently passing in front of (or through) the
southern portion of the Pleiades \citep{gordon84}
and this cloud causes several of the low mass Pleiades stars to
be relatively heavily reddened.  Figures \ref{fig:jonesbvv},
\ref{fig:otherbvv}a and \ref{fig:otherbvv}b
\citep[and Figure 1 of][]{jones72} assumed that all Pleiades
members have the same (small) reddening, i.e., $E(B-V)\,=\,0.04$. 
Several of the stars that lie above the main Pleiades locus in those
figures are stars that are located behind the molecular cloud and
are heavily reddened.\footnote{
The published reddening estimates for these Pleiades
K dwarfs were based on assuming these stars have intrinsic $B-V$
colors that are ``normal'' for their spectral type.  We cannot assume
that any longer.   Therefore, we have rederived reddening estimates
for these stars.  The new estimates assume that these Pleiades
K dwarfs have $V-I$ colors that are normal for their spectral type.
The rationale for that choice is based on the analysis presented
later in this section.   Our qualitative results would have been the
same if we had simply retained the published estimates. }
Figure \ref{fig:otherbvv}d provides a new Pleiades CMD, where we have
1) combined data from all of the photometric surveys and 2) individually
de-reddened the stars believed to be affected by the extra reddening
from the molecular cloud. The new figure has significantly fewer stars
displaced above the ZAMS curve for $B-V > 0.9$.  Of the eight Pleiades K
dwarfs above the ZAMS curve for $B-V > 0.9$ in Figure \ref{fig:otherbvv}d,
five (HII 885, 3197, 1100, 1553 and 1348) are known binary or
triple systems with mass ratios near one 
\citep{bouvier97,queloz98}. If approximately corrected for their binarity
by subtracting 0.7 mag from their $\mbox{M}_{V}$, those stars would
fall well below the ZAMS curve and within the locus of the other Pleiades
K dwarfs.  We therefore conclude that the mechanism affecting the Pleiades
K dwarfs is ubiquitous.

A more direct way, perhaps, to illustrate the Pleiades
anomaly is to compare the Pleiades photometry for
low mass stars to data in another open cluster.  The best
cluster for that purpose is Praesepe.  Praesepe is very
similar to the Pleiades in terms of richness and distance,
but is 5-10 times older.  If the Pleiades anomaly is
age-related, it is therefore plausible that the Praesepe
stars would be unaffected.   Furthermore, photometry of 
Praesepe and the Pleiades exists which was obtained with
the same telescopes and photometers and utilizing many of the
same calibration stars
\citep{upgren79, weis81, stauffer80, stauffer82a, stauffer82b, stauffer84},
minimizing the possibility of systematic errors in the
calibrated colors.
For this reason we will use only these sets of photometry in the next figure.
The one difficulty with the use of Praesepe for this purpose
is that Praesepe may be slightly metal rich relative to the
Sun, whereas the Pleiades is solar metallicity within the errors
(see Appendix \ref{app:metals} for a brief review of spectroscopic
and photometric metallicity estimates for the two clusters).
In the plots to follow, we adopt MS fitting distance moduli
for the two clusters chosen to align the single
star main sequence loci of the two clusters for F and G
dwarfs, thus subsuming to first order the metallicity
shift into the distance moduli.
\footnote{
    Since we are comparing the shapes of the main
    sequences in the two clusters, this procedure should
    be adequate.  Theoretical models predict very
    little differential shift in the shape of the
    main sequence loci as a function of metallicity for
    the small metallicity shift between the two clusters.
    Using the Boessgaard \& Friel metallicity scale, the
    more metal-rich Praesepe main sequence should be
    $\sim 0.1$ mag fainter in $M_V$ at constant
    $B - V$ \citep[e.g.,][]{pinsono98}, so Praesepe
    is actually correspondingly more distant than the
    apparent distance modulus we adopted.}  
The values we selected for the Pleiades are $(m-M)_0 = 5.53$,
$A_V = 0.12$ \citep{becker71,jones72,crawford76,turner79,nissen88};
while for Praesepe we chose $(m-M)_0 = 6.05$, $A_V = 0.0$
\citep{crawford69,becker71,nissen88}.
Figure \ref{fig:plecmds}a compares photometry for the
Pleiades low mass stars to low mass stars in Praesepe
in a color-magnitude diagram again using $B-V$ for the
abscissa. The diagram shows the same displacement of
the Pleiades K dwarfs below the ZAMS (as defined by the
Praesepe stars) as illustrated in Figure \ref{fig:jonesbvv}.

If the anomaly affecting the Pleiades K dwarfs causes them
to be dimmed at $V$ but does not significantly alter their
colors (as posited by Jones), then one should see the same
basic effect in a color-magnitude diagram using a different
broad-band color for the x-axis.  The above set of authors
also obtained $V-I$ colors for Pleiades and Praesepe
low mass stars, and therefore one can directly test the
veracity of this prediction.  Figure \ref{fig:plecmds}b
provides such a diagram for the two clusters.  This diagram
tells a much different tale: the Pleiades and Praesepe 
low-mass stars trace coincident loci for the entire magnitude
range for which Kron-system $V-I$ phototube photometry is
available.\footnote{
Note that the Kron-system $V-I$ photometry used here differs
from the more commonly used Cousins-system with
$(V-I)_{\mbox{K}}\,\sim\,(V-I)_{\mbox{C}}\,-\,0.2\,\mbox{}$
for early to mid-K dwarfs. See \citet{bessellweis} for a
more accurate conversion formula.}

The results for the $V-I$ color-magnitude diagram are
strongly contradictory to the neutral-extinction dust
shell hypothesis.  Instead, it appears that it is more
likely that the correct diagnosis of the anomaly affecting
the Pleiades K dwarfs must rely on a mechanism which
produces more flux in $B$ without significantly
altering $V$ or $I$.

Because the two CMD's discussed so far show fundamentally
different behavior, it is useful to construct a third
diagram.  Fortunately, the 2MASS all-sky near-infrared
survey \citep{skrutskie97} provides accurate and homogeneous
$JHK_S$ photometry for most of the stars shown in 
Figures \ref{fig:plecmds}ab.  Thus, Figure \ref{fig:plecmds}c
provides a comparison of the two clusters in an
$M_{V}$ versus $V-K_S$ diagram.  Somewhat disconcertingly,
this diagram shows a fundamentally different character from
the preceding two.  In this case, the Pleiades late K dwarfs
and early M dwarfs fainter than $M_V \sim 8.5$ are displaced
systematically above the Praesepe main sequence locus.

If the last panel of Figure \ref{fig:plecmds} were the only
diagram under consideration, its interpretation would be
qualitatively straightforward.  That is, the low-mass stars
in the younger cluster fall above the main-sequence locus
defined by the older cluster because the younger stars are
still contracting to the main sequence.  Comparison to the
\citet{baraffe98} isochrones, using both the absolute $V$
magnitude for the PMS turnon point and the displacement of
the lower mass stars above the ZAMS, suggests an age of
order 70-100 Myr for the Pleiades, which is not an unreasonable
value when compared to recent age estimates from other
methods \citep{ventura98, ssk98}.  However, we do
have two other diagrams, one of which shows the Pleiades
low-mass stars below the Praesepe locus and the other
showing the two loci to be coincident to $M_V \sim 10.5$.
The simple empirical conclusion is that the derivation
of accurate PMS isochrone-fitting age for the Pleiades
depends on the choice of colors in the CMD and therefore
is not at all straightforward.

\section{Spectroscopic Confirmation of the Pleiades Color Anomaly}
\label{sec:spectra}

In order to provide further constraints on the physical
mechanism causing the anomalous colors for the Pleiades K
dwarfs, we have obtained intermediate-resolution spectra
of a small sample of Pleiades and Praesepe stars using the
Low Resolution Imaging Spectrograph (LRIS) and the Keck I
telescope on Nov.\ 21, 2001.  For a description of LRIS,
see \citet{oke95}.  The stars observed at Keck are
documented in Table \ref{tab:lris}.

Each of the stars was observed with the blue and red
channels of LRIS using a 400 lines mm$^{-1}$ grating
which provided a spectral resolution of about 6.0 \AA.
The blue spectra covered $3300-5300$ \AA\ and the red
spectra covered $5700-8400$ \AA.  All of the spectra were
obtained at airmass less than 1.015 and were extracted
and flux calibrated in the same manner.  The spectra were
taken on a night of variable cloud cover, but we believe
that these observations were taken when few or no clouds
were present.  In any event, our arguments do not depend
on the absolute flux calibration, but only on the relative
shapes of the individual spectra.

For this analysis, we chose two Pleiades K dwarfs
whose $B-V$ colors are particularly blue (as indicated in
the $M_V$ vs. $B-V$ diagram), and then selected comparison
Praesepe K dwarfs with as nearly the same $V-I$ color
as possible.  
For each pair, we have scaled the Praesepe star's flux (at
5300 \AA\ for the blue spectra and 6000 \AA\ for the red
spectra) to that of its Pleiades counterpart. 
Figure \ref{fig:lris} shows the individual blue spectra
for one of the pairs.
The only gross difference that is readily evident is
that the Pleiades member (HII~2927) has \ion{Ca}{2} H+K
in emission, whereas for the Praesepe member (KW~575)
those lines are in absorption.  This is, of course,
expected given the correlation between age, rotation
and chromospheric activity for low mass stars.

In order to illustrate
more easily that there is a significant difference in the
spectral energy distribution of the two stars, we 
divided the Pleiades spectrum by the scaled version of
the Praesepe spectrum.  The two ratio spectra are shown in
Figure \ref{fig:normblue} (for the blue spectra) and
Figure \ref{fig:normred} (for the red spectra). 
The ratio spectra have very similar appearances in both cases.
In the blue, both spectra show a blue continuum,
rising to shorter wavelength and both show
residual emission lines for several members of the Balmer
series as well as \ion{Ca}{2} H+K.  In addition, both ratio
spectra show two other identifiable quasi-emission features
- \ion{Ca}{1} 4226 \AA\ and MgH 4780 \AA, indicating that
these features are weaker in the Pleiades stars than their
Praesepe counterparts.  Both these features are strongly
temperature sensitive in this spectral range, increasing
in strength for later spectral types.  By integrating
over the $B$ band for the ratio spectra, we find that
the Pleiades K dwarfs are about 10\% brighter than their
Praesepe counterparts (when scaled at 5300 \AA) - and
hence should be about 0.1 mag bluer in $B-V$, in agreement
with the broadband photometry.   For the red channel LRIS
data, the ratio spectra show H$\alpha$\ in emission,
a quasi-emission line at \ion{Na}{1} 5893 \AA, and
well-defined negative residuals at the TiO band locations.
That is, the inferred spectral type for the Pleiades stars
would be earlier than their Praesepe counterpart in the
blue, but later than their Praesepe counterpart in the red.

To zeroth order, the spectroscopy simply serves to confirm
that the photometric color anomaly is real.  We can also
say that while the Balmer and CaII H+K emission contribute
to the blue excess, the dominant cause for the bluer $B-V$
color is the enhanced blue continuum.

\section{Cool Spots and Young K Dwarfs}
\label{sec:spots}

\citet{vleeuwen86} and vLAM87 obtained a
large body of accurate photometry of low mass stars
in the Pleiades in the Walraven intermediate band system. 
Those authors made a fundamentally important contribution to
the present topic by obtaining multiple observations of many
of their targets.  In particular, they determined that
at least $10\%$ of the Pleiades K dwarfs have periodic
photometric variability with periods of order 6-24
hours and $V$ magnitude amplitudes up to $\sim$0.15 mag.
No periods were derived for Pleiades G dwarfs and, on
average, the G dwarfs are significantly less variable
than the K dwarfs.  High resolution spectroscopic studies
\citep[e.g.,][]{sh87,queloz98,terndrup00} have since shown
that there is indeed a wide range in rotational velocity
amongst the Pleiades K dwarfs, and that the more slowly
rotating K dwarfs generally are less photometrically
variable.

A number of other groups have subsequently derived
additional light curves and periods for low mass Pleiades
members \citep{shs84, prosser93, prosser95, krishna98}.
Figure \ref{fig:rotvar} provides another $M_V$ versus
$B-V$ CMD for the Pleiades, but this time the shortest
period and/or most variable Pleiades K dwarfs are highlighted.
It is apparent that there is a good correlation between
rapid rotation, or the large amplitude variability that
normally accompanies rapid rotation, and the amount of
blueing in $B-V$.  Note that many of the stars in this
figure do not have known rotation periods, so the fact
that a star appears only as a small filled dot does not 
necessarily mean it is a slow rotator.  However, about 
2/3 of the Pleiades K dwarfs are known to have relatively 
small $\mbox{v} \sin i$ ($< 20$ \kms), so most of the small 
filled dots are likely to be relatively slowly rotating 
and probably less spotted.

Another correlation between variability and the abnormal
SED's is illustrated in Figure \ref{fig:deltav}, which shows
the difference between the \citet{jm58} $V$ magnitude and the
Stauffer $V$ magnitude as a function of luminosity.  One would
expect the difference to increase for fainter stars simply
due to photon statistics, but the dispersion shown for $V >
12$ is far larger than expected from measurements errors.
In fact, \citet{jm58}, had originally noted this increased
variability fainter than V = 12 based on the stars for which
they had more than one measurement.
The $V$ magnitude where the level of variability sharply
increases ($V \sim 12$) is the same as where the anomalous
$B-V$ colors become evident.
As noted by vLAM87 and \citet{stauffer84}, the catalogs of
flare stars in the Pleiades also indicates that
there is a sharp transition to most cluster members being
flare stars at V $\sim$ 12.5 (with no Pleiades member being
identified as a flare star brighter than V $\sim$\ 12.0).

The preceding two figures demonstrate that there is at
least a correlation between spottedness and the Pleiades
K dwarf color anomaly.  Is it possible to explain the
color anomaly simply as the direct result of laterally
inhomogeneous effective temperature distributions (i.e.,
spots) on the Pleiades K dwarfs?  If the only process
involved were the formation of cool spots and then a
concurrent slight increase in radius of that star (to allow
the luminosity to remain constant over the long term),
then cool spots could not produce the observed effect
since the resultant star would lie above a main-sequence
locus defined by stars without such spots.  However, if
one allows both cool and warm spots (i.e., plage areas),
or if one simply allows the quiescent photosphere to be
warmer than it would be if it were not spotted,
then in principle one could account for the observed
color anomaly.

As a proof of concept, we have attempted to match the
$BVIJHK$ spectral energy distribution for representative
mid-K dwarfs in the Pleiades by a linear combination
of the spectral energy distribution of (assumed to be)
unspotted dwarfs. We assume that the Pleiades K dwarfs
have photospheres composed primarily of two characteristic
temperatures - a warm component and a cool component.
We selected a sample of approximately a dozen field K
dwarfs with $BVRI$ photometry from \citet{bessell90} to
represent the former temperature range, and a similar
set of M dwarfs from the same reference to represent
the cooler component.  We extracted $JHK$ photometry for
these stars from the 2MASS database.   For the Pleiades
stars, we used the $BVRI$ photometry of Stauffer from the
1980's, adopting average values for the stars when there
are multiple observations.  Because the optical and IR
photometry were obtained at widely different epochs,
and the stars have $V$ magnitudes that vary by up to 0.2
mag, we allow for a zero point offset between the $BVRI$
and the $JHK$ photometry of up to 15\%.   We created
two-component models for all pairs of field K and M dwarfs
in our sample, and for a wide range in the relative $V$ band
luminosity ratios for the two components.  We normalized
the model SED's to the Pleiades K dwarf $V$ magnitude,
and then searched for the model SED (or SED's) which
deviated least from the Pleiades star SED in an RMS sense.
We doubly weighted the $B$ band point because the goal
of the exercise was to see if these simple spot models
could explain the $M_V$ vs. $B-V$ CMD for the Pleiades.
The best fits for four of the Pleiades K dwarfs are described
in Table \ref{tab:models}.  On average, the calculations
indicate that the best fits come for models with of order 
$15\%$ of the $V$ band light coming from a cool component
with an early M dwarf spectral type, and the remaining
flux coming from an early K dwarf.  With these V flux 
ratios and spectral inferences, the fraction of the 
surface area of the Pleiades K dwarfs covered by
``cool spots'' would be of order $\gtrsim$ 50\%.

The very large areal coverage implied by our spotted star SED
models is not necessarily in contradiction to the conclusions
drawn from attempting to model the light curves for the spotted
stars \citep[e.g.,][]{sda91,bouvier96}, where spot covering
fractions of order 15\% are normally derived, because the 
light-curve models are only sensitive to the 
non-axisymmetrically distributed spots, whereas 
the SED model is sensitive to spots even if they
are distributed homogeneously over the surface of the star.

We further tested the hypothesis that the late-K stars
in the Pleiades have photospheres with regions of different 
temperature by comparing the LRIS spectra of the stars in
Table \ref{tab:lris} with spectra in the library
assembled by \citet[][hereafter JHC]{jhc84}. 
The library consists of flux-calibrated spectra covering
3510 -- 7427 \AA\ at a resolution of $\sim 4.5$ \AA, and
are therefore well matched to our Keck spectra.

First, we systematically compared our LRIS spectra to each
of the JHC spectra, which in their library are sorted by subtype.
For each spectrum $i$ in the library, we computed the
r.m.s.- difference $\sigma(i)$ between the library spectrum 
and a Keck spectrum, where both were scaled to have the
same average flux.  The spectral type was taken as
the index $i$ that produced the minimum value of $\sigma(i)$.
A fractional spectral type was found by fitting a parabola
to $\sigma(i)$ and using the fit minimum as the spectral type.
The comparison was done with all spectra smoothed
to a resolution of 5 \AA\ and the Pleiades spectra
corrected to zero reddening (the JHC spectra themselves
are similarly corrected) using the \citet{rk85} reddening law.
What we are doing with this exercise it to compare
the overall shape of the SED for each star against the
JHC library rather than comparing absorption-line strengths.

The determination of the spectral type was done separately
for the blue and red portions of the Keck spectra with
the results compared in Figure \ref{fig:spts}. 
Each axis shows the index $i$ for the JHC spectra of
late-type main-sequence stars (the full library has
161 spectra of several luminosity classes); 
the corresponding spectral classes are shown on the
top and right axes.  Filled points show the spectral
classification for the Praesepe stars, while open points
are for the Pleiades.  The solid line indicates equality
between the two and is not a fit to the data.  
We find that the Praesepe stars all have equal spectral
types in the blue and red regions of the spectrum,
meaning that they have the same overall SED as the
stars in the JHC library.  The Pleiades stars, on
the other hand, have a spectral type in the blue that
is earlier than that found for the red, consistent
with our pairwise matching of Pleiades and Praesepe
spectra above.

We then proceeded to model the individual spectra with a
two-component fit, finding the best linear combination
of pairs of stars in the JHC library that matched
our Keck spectra.  For the Praesepe stars, the fit
was not improved by this process, meaning that a
two-component fit did no better than a fit to
a single spectrum.  The Pleiades spectra, on the
other hand, were usually fit best by a combination
of early K and early M dwarf spectra (e.g., K2 + M1),
consistent with our analysis of the broad-band colors above. 
When compared to the best fit composite spectra, the Pleiades
stars had nearly the same overall SED, but still exhibited
positive residuals in \ion{Ca}{2} H+K as was the case
when compared to the Praesepe spectra.  Thus we
conclude the Pleiades stars have more than one
photospheric temperature whether we compare the colors
or by comparison to a spectral library.

\section{Evidence for this Effect Beyond the Pleiades}
\label{sec:beyond}

If the anomalous spectral energy distribution found in
the Pleiades K dwarfs is indeed linked to the youth of the
cluster, then the same spectral energy anomaly should be
present in other young K dwarfs.  Because the Praesepe K
dwarfs are not affected, we can assume that by ages of
order 600 Myr the condition has dissipated.   Is there
existing observational data to support these conclusions,
and are we able to define better the age range where there
is a significant effect?

At very young ages - a few million years or less - the
combination of variable extinction, accretion disks, hot
spots from accretion columns, etc. make the detection
of the relatively subtle effect under consideration
unfavorable.  However, by isolating weak-lined T Tauri
stars (WTT's), both because they may be systematically
older and they lack significant circumstellar disk
signatures, it is possible to search for the $B-V$ anomaly.
Because in some cases we do not know distances, and
because we have to combine data for multiple stellar
regions, the best way to search for the effect is in
a two-color diagram.   We have searched the \citet{kh95}
database for WTT's with $BVI$ photometry,
and plot those stars in a $B-V$ vs. $V-I$ diagram as
Figure \ref{fig:wtts}, where we compare the location of
the WTT's to Pleiades and Praesepe dwarfs.   
The figure shows that, indeed, the WTT's are blue in $B-V$
for their $V-I$ colors, and preferentially track the location
of the Pleiades stars in the diagram. Further, 
some of these Taurus WTTs have been shown to have
anomolous colors both at $B-V$ and at other red optical and
optical-infrared colors by \citet{gullbring98}, who also
suggested that large areal spots were the probable cause.
We conclude that the Pleiades anomaly is present at a
few Myr, and that the spectral type range affected
is similar (with perhaps a hint that the effect
extends to earlier spectral type at the younger age).

The open cluster NGC~2547 is younger than the Pleiades, having an
age between 15 and 60 Myr \citep{claria82,jeffries98,naylor02}.
Probable cluster members were identified by \citet{jeffries98} 
using X-ray activity and the location of the sources
relative to the zero-age main sequence. In Figure \ref{fig:n2547}
we use the photometry and candidacy criteria of \citeauthor{jeffries98}
to plot the color-magnitude diagram(s) for NGC~2547, again
employing both $B-V$ and $V-I$ as temperature surrogates.
In these figures we also plot the sources relative to 
theoretical pre-main sequence isochrones taken from 
\citet{dm97} and converted to observable quantities ($BVI$)
following the methods described in \citet{shb95,stauffer97}
and detailed in Appendix \ref{app:convert}.
Briefly, these conversions were tuned to require the Pleiades
lower main sequence to lie along the 100 Myr \citeauthor{dm97}
isochrone and to fit the observed $B-V$ and $V-I$ colors
of nearby field dwarfs. Similar to the behavior of the
Pleiades in Figure \ref{fig:plecmds}, the two CMDs tell
somewhat different stories. Using $B-V$ on the X axis, we
find that the G-type members are coincident with the
main sequence, while the K dwarfs have an inferred contraction age
of 35-40 Myr. Using $V-I$, on the other hand, suggests that
the G stars lie well above the \citeauthor{ji58} MS,
and that the inferred age of the K dwarfs is 25-30 Myr.
This 5-10 Myr difference between the ages inferred from
the CMDs in Figure(s) \ref{fig:n2547}ab was also found by
\citet{naylor02}, and these examinations of NGC~2547
support our conclusions from studying WTTs that the Pleiades
anomaly extends to younger ages, where it further pollutes
earlier spectral types probably up to at least G.

It is difficult to infer the timescale over which the Pleiades
anomaly might disappear from the K dwarfs because few open clusters
intermediate to the Pleiades and the Praesepe have sufficiently
accurate membership criteria to confirm that the Pleiades anomaly 
is still present in any of them.  A typical example is NGC~2516,
which has an age $\sim\,140$ Myr \citep{meynet93}.  Two recent papers
have published extensive photometry in $BVI$ for this cluster
\citep{jeffries01,sung02}.  The metallicity and distance
modulus for NGC~2516 is a subject of some controversy, but
now appears to be near solar according to \citet[][and references
therein]{terndrup02}, who find $(m-M) = 8.49$ and
$E(B-V) = 0.13$.  Figure \ref{fig:n2516} provides the
$B-V$ based CMD for NGC~2516 with these assumptions and using
the Jeffries et al.\ photometry. Clearly, the locus of
points appears to follow the \citeauthor{ji58} MS fairly
well, however there may be a hint of possible K dwarf members
scattering to bluer $B-V$.

\section{Claims for an Age Spread in the Pleiades - Real
or Illusory?}

As noted in the introduction, \citet{herbig62} used a $B-V$
based CMD for the Pleiades in his landmark paper.  He found
the low mass Pleiades stars to be essentially on the main
sequence to spectral type K5 or $M_V \sim 8.5$.  Because
there were as yet no detailed PMS theoretical evolution
models available to him, Herbig used the Kelvin-Helmholtz 
(KH) formula
to estimate the age corresponding to a given PMS turnon
point.  For $M_V = 8.5$, he derived a contraction age of
220 Myr.  Because that was much older than the nuclear age
of $\sim$ 60 Myr that he adopted, he concluded that there
was a large age spread in the Pleiades with the low mass
stars being significantly older on average than the high
mass stars.

In retrospect, three factors contributed to Herbig getting
what we believe to be the wrong answer.  First, the main
sequence he used was somewhat inaccurate, dipping about
0.4 mag below the \citeauthor{ji58} ZAMS at $B-V = 1.3$.
Had he adopted the \citeauthor{ji58} main sequence, he
would have faced the same quandary as \citet{jones72}, and
perhaps concluded that no PMS age derivation was possible
(because the low mass cluster members are below the main
sequence, rather than above).  Second, the anomalous $B-V$
colors of the Pleiades K dwarfs clearly had an influence:
if these stars had ``normal'' $B-V$ colors, then they
would have fallen above the ZAMS Herbig used, and he
would have at least derived a contraction age younger
than 220 Myr.  However, the dominant factor which caused
Herbig to derive too old an age for the Pleiades was the
use of the Kelvin-Helmholtz formula to convert the
PMS turnon point to an age.  Herbig noted that the KH
formula predicted a contraction age of 60 Myr for solar
mass, and 250 Myr for $M_V = 8.5$.  These ages are about
twice as old as what would be inferred from modern PMS
evolutionary models.  For example, if Herbig had used the
\citet{baraffe98} tracks, for $M_V = 8.5$, he would
have derived an age of $\sim$ 100 Myr, much closer to the
adopted nuclear age.

More modern claims of significant age spreads in the
Pleiades (as well as other clusters) have continued to be
cited as evidence for bimodal star formation.
The two most recent claims for a large age spread in the
Pleiades can be found in \citet{siess97} and \citet{belikov98}. 
\citet{siess97} used the width of the stellar distribution for
K and M dwarfs in an $M_V$ versus $B-V$ CMD to estimate an
age spread for the Pleiades low mass stars of about 30 Myr.
\citet{belikov98} converted an $M_V$ versus $B-V$ CMD to
a luminosity function, and inferred an age spread of 20 to
60 Myr for the low mass stars in the Pleiades by comparison
of the detailed structure of the LF to a theoretical
model (a bump in the LF is expected where the PMS track
joins the MS).  Analysis of the Pleiades photometry
indicates to us that neither of these age indicators can
be trusted.  The Siess et al.\ analysis is predicated on
the assumption that mechanisms other than an age spread
do not introduce a significant dispersion to the $B-V$
based CMD in the absolute magnitude range $6 < M_V < 10$.
However, it is clear from examination of our CMD's that
the spot-related mechanism affecting the spectral-energy
distribution of the Pleiades K dwarfs (as well as simply
their variability) does introduce a significant spread in
the CMD.  With respect to the Belikov et al. age spread,
the bump in the LF that they attribute to the PMS turnon
point in the Pleiades occurs at $M_V = 5.5$. 
Our Figures \ref{fig:plecmds}bc show that the PMS turnon
point in the Pleiades is almost certainly fainter than
$M_V = 8.5$, and that nothing related to a real PMS turnon
point is present in the $B-V$ based CMD near $M_V = 5.5$.

\section{Discussion and Speculation}
\label{sec:discuss}

While it is tempting to ascribe a causal relationship
between the presence of spots and the anomalous colors for
the Pleiades K dwarfs, it is possible that both are merely
products of some other process.  The fundamental physical
phenomenon that is is probably driving both observables is
the large amount of non-radiative heating being deposited
into the outer layers of these stars due to their rapid
(differential) rotation.  One quantitative measure of
the amount of non-radiative heating is the X-ray flux.
Figure \ref{fig:xray} shows the relation between X-ray
flux and $B-V$ color for the Pleiades as derived from ROSAT 
and updated from \citet[][Micela, private communication]{micela99}.
The figure indicates that the mean fractional luminosity
emitted in X-rays rises steeply through the F and G spectral
range, and peaks at about mid-K ($B-V$ = 1.0), and then
stays constant through early M.  For the K stars, the X-ray 
flux accounts for 0.1\% of the luminosity of the star.  
The coronae also emit X-ray photons downward, so a similar 
amount of energy is also deposited in the upper photosphere 
and reprocessed.  Since the excess $B$ band flux (relative 
to older dwarfs) for the Pleiades K stars is of order 
0.4\% of the star's luminosity, modifications to the 
optical SED of these stars due to reprocessing of coronal 
flux could, in principle, account for a significant part 
of the blue excess.

One additional clue to a physical connection between
the anomalous SEDs of the Pleiades K dwarfs, their coronal
and chromospheric activity and the star's rapid rotation is
presented in Figure \ref{fig:vsini}. Here we plot the degree
of the distortion of the K dwarf SEDs, calculated as difference
of the Pleiades dwarf's V band magnitude and that of a
main sequence dwarf at constant $B-V$, as a function of
the star's  $\mbox{v}\,\sin{i}$. In addition to showing that
all the rapid rotators show the anomaly, this figure clearly 
indicates that larger $\mbox{v}\,\sin{i}$ do not yield systematically
larger 'blueing' of the dwarf's SED; there is a maximum
amount of blueing possible even at large rotation rates.
The fact that coronal and chromospheric activity in young stars
display similar ``saturation'' relationships \citep{stauffer94}
suggests that all three (spottedness, coronal activity and the
anomolous SEDs) are all driven by rotation, and possibly that
the photospheres of these stars reaches a maximum level of
spottedness that cannot be exceeded. Our analysis suggests
though that this degree of spottedness can be quite large,
of order $\sim50\%$.

As noted in the introduction, vLAM87 provided evidence for a blue
or ultraviolet excess (relative to inactive field stars) for a
sample of Pleiades K dwarfs.  Those authors speculated that the
excess was due to chromospheric emission rather than spots. 
While it must be true that chromospheric emission contributes
to the anomalous colors of the Pleiades K dwarfs, we do not 
believe it is the dominant mechanism because (a) it would not
explain the wavelength dependent spectral types
(for $\lambda$ $>$\ 4000\AA) in the Pleiades stars relative to
old field K dwarfs; and (b) the ratio spectra shown in
Figure \ref{fig:normblue} do not show evidence of any change at
the Balmer jump ($\lambda$\ $\sim$ 3650 \AA), as predicted by
vLAM87.  Also, as noted above, we cannot decompose
the spectrum into a ``normal'' spectrum plus a smooth continuum
\citep[as in classical T-Tauri stars, e.g., ][]{hartigan91},
showing that the excess is not due to Paschen and free-free
continuum emission. Thus, our usage of the term ``spot'' is
meant to include both dark (cool) spots and plage areas 
{\em on the photosphere} and is, therefore, distinct from
``chromospheric emission,'' which we define as emission from
regions above the temperature minimum, not the photosphere.

Regardless of the cause of the SED anomaly, the fact that
SED of the Pleiades K dwarfs is significantly different
from ``normal'' means that the effective temperature
one estimates for a Pleiades K dwarf is ill-defined.
The 0.1 mag difference in $B-V$ corresponds to about a
250 K difference in T$_{eff}$ - or, alternatively, the T$_{eff}$
estimated from $B-V$ would differ by about 250 K from the
T$_{eff}$ estimated from $V-I$, and even more different from
a T$_{eff}$ estimated from $V-K$.  This, by itself, would
indirectly affect (for example) metallicity estimates.
Because the integrated light from these stars originates
from regions with a wide range of effective temperatures,
individual line strengths and ratios will also differ from
``normal'', more directly affecting metallicity estimates.
The Pleiades K dwarfs with color anomalies are also the
subject of a long-standing controversy regarding their
apparent lithium abundance spread \citep{king00a}.
Spots have been considered as an explanation for that
spread in the past \citep[e.g.,][]{soderblom93},
however that explanation has normally been
dismissed because the degree of spottedness needed
seemed too high \citep{dbn01a}.
Our photometry and spectroscopy indicates that
the cool component may be contributing 20\% or more
of the light for the Pleiades K stars, possibly enough to
explain at least most of the observed lithium spread.

The distance modulus for the Pleiades that we have used in our
figures is the main-sequence fitting distance, rather than the
Hipparcos distance \citep{vleeuwen99}, which is significantly
different. A number of papers have argued that the Hipparcos
distance to the Pleiades and some other open clusters is  
inaccurate due to correlated errors on small (1 degree) angular
scales in the Hipparcos astrometry
\citep{pinsono98, naray99b,makarov02}.
Other authors \citep{lindegren00,vleeuwen97} have argued
that those correlations are not large enough to align the Hipparcos
and main-sequence fitting distances.  Alternate explanations that
have been advanced include metallicity or helium abundance
values that differ significantly from standard values for these clusters or
an unknown age-related process \citep{vleeuwen99,robichon99,grenon01}. 
Our results might seem to offer support for the latter position.
We do not believe that this is the case.
The blue-excess for the K dwarfs would cause the main-sequence
fitting technique to derive a too large distance for the
Pleiades, which is in fact the sense of the difference
with Hipparcos.  However, this is not, in fact, explanatory
of the discrepancy between the MS-fitting and Hipparcos
distances for the Pleiades because most of the distance
moduli quoted in Section \ref{sec:archival} were based on
observations of stars with~$\mass\,>\,1\,\solarmass$, whereas
the ``K dwarf anomaly'' is only evident for
$\mass\,<\,0.8\,\solarmass$.  For the \citeauthor{pinsono98}
paper, the $B-V$ based CMD utilized was truncated at
$B-V\,=\,1.0$ in order to avoid this problem.
Also, \citeauthor{pinsono98} used CMD's with $V-I$ as the
color-index as well, where the spot influence is apparently minimal.  
Nevertheless, small systematic color anomalies can cause large
distance modulus errors due to the relatively steep slope of the MS,
so this problem needs to be kept in mind.

Finally, we note that it should be possible to utilize
the anomalous $B-V$ colors of young, field K dwarfs as a
means to identify them in large-scale photometric surveys.
That is, if an all-sky $BVRI$ photometric survey (or some
similar dataset) ever becomes available, one could use a
$B-V$ vs. $V-I$ diagram to select probable young K dwarfs.
Or, if a future space mission obtains accurate photometry
and parallaxes for a complete set of stars fainter than
Hipparcos, one could use a $B-V$ based CMD to select
possible young K dwarfs with followup spectroscopy to sort
the young K dwarfs from the low metallicity contaminants.
In fact, even our Figure \ref{fig:hippbvv} (from Hipparcos)
may harbor a few young K dwarfs amongst the field stars
with $1.0 < B-V < 1.3$ and with $M_V$ that fall up to
about 0.5 magnitudes below the main sequence curve. 
We do not expect many such young K dwarfs because the volume
of space sampled is small, but a few would be possible. 
A literature search does indeed find three stars located in
the appropriate part of Figure \ref{fig:hippbvv} that may
be counterparts to the Pleiades K dwarfs: 
Gl~174 (=V834Tau), $M_V =7.44$, $B-V = 1.09$;
Gl~517, $M_V = 7.76$, $B-V = 1.21$; and
GJ~1257 $M_V = 8.02$, $B-V = 1.12$). 
All three appear to be young based on chromospheric or
coronal emission, flaring or variability, and/or space motions.

\section{Summary and Implications}
\label{sec:summary}

K dwarfs in the Pleiades have spectral energy distributions
that differ from relatively old K dwarfs.  The primary
difference is that the Pleiades stars have blue excesses
(bluer $B-V$ colors), but they also probably have slight
near-IR excesses also.  The effect is greatest for rapid
rotators.   The color anomaly is well correlated with
spottedness as inferred from photometric light curves,
and may largely be explained by spots.  If so, however,
the fractional area of the surface of these stars covered
by cool spots must be very large ($>$ 50\%).

We believe that these anomalous SED's are likely to be
present for all young K dwarfs, being largest for the
youngest and most rapidly rotating stars.  Praesepe members
do not appear to share this anomaly, and therefore the
upper age limit for the effect is of order 600 Myr.
As previously noted by \citep{gullbring98}, this effect will
result in young K dwarfs having inferred spectral types that
are a function of wavelength with the apparent type being
later at longer wavelengths.

One negative impact of this anomaly is that we believe it
becomes much more difficult to derive a pre-main sequence
isochrone age estimate for clusters in the 50 $<$ age $<$
200 Myr age range, because the displacement of the low
mass stars above the ZAMS in an L-T$_{eff}$ diagram becomes
comparable to the uncertainty in placing the stars into the
theoretical plane. Additionally, this uncertainty means 
that observational tests of theoretical pre-main
sequence tracks (e.g., coeval binaries, cluster isochrone
fitting) will be limited in the precision they can provide.
Detailed spectral analysis of young
K dwarfs to derive [Fe/H] or other elemental abundances
(e.g., lithium) also becomes much more uncertain if the
degree of lateral temperature inhomogeneities is as large
as we conclude.

A positive impact of the realization that this color
anomaly is present is that it should be possible to
identify young K dwarfs simply from broad-band photometry.
A few such stars may be identifiable already in the
Hipparcos database \citep{hipparcos}; many more could be
identified with GAIA \citep{lindegren96}.

\acknowledgments

The authors thank Giusi Micela for the use of Figure
\ref{fig:xray} in advance of publication.


\appendix

\section{Metallicity of the Pleiades and the Praeasepe}
\label{app:metals}

Surprisingly, there have been relatively few determinations of
metallicity in the Pleiades and Praesepe.  Because we are primarily
interested in knowing whether there is a difference in metallicity
between the two clusters, we prefer to restrict our consideration
to studies where we can compare results for both clusters derived
from similar datasets and using as similar analysis methods as
possible.  There are three such pairs of studies using spectroscopic data: 
(a) \citet{boesgaard89} found Praesepe 0.07 dex more metal rich than
the Pleiades; 
(b) \citet{fb92} and \citet{boesgaard90} found Praesepe to be more
metal rich by 0.072 dex; and
(c) \citet{boesgaard88a} and \citet{boesgaard88b} found Praesepe
to be more metal rich by 0.13 dex.  
The most recent spectroscopic determinations used up to
15 Fe lines between 6591 and 7832 \AA\ and included 12 Pleiades
and 6 Praesepe targets.  There are also relatively
recent studies of both clusters using Stromgren photometry; they
have the result that Praesepe is found to be more metal rich than
the Pleiades by 0.13 dex.
In Table \ref{tab:app:metals}, we list the individual [Fe/H]
estimates for both clusters and errors quoted from the references
themselves.

\section{Areal Coverage of Cool Spots for the Pleiades K dwarfs}
\label{app:sedfit}

We use the results of our empirical fits to the spectral energy distribution
(SED) of the Pleiades K dwarf HII~3063 (see Table \ref{tab:lris} for magnitudes 
and colors) to derive the areal coverage of ``cool spots.''  

Our best fit two component model of the SED of HII~3063 (as described in
Section \ref{sec:spots} and listed in Table \ref{tab:models}) consisted of
a warm component whose SED is assumed to have the same shape as the field
K5~dwarf Gl~3 ($B-V\,=\,1.06;\;V-I_{C}\,=\,1.20$), and a cool component 
whose SED is assumed to have the same shape as the M2~dwarf Gl~382
($B-V\,=\,1.52;\;V-I_{C}\,=\,2.19$).
Using a conversion scale from \citet{bessell79}, the field dwarf
$V-I_{C}$~color of Gl~3 corresponds to a surface temperature of
$4560\,\mbox{K}$.
For the Gl~382 M~dwarf, we use the $V-I_{C}\mbox{ to T}_{eff}$ relationship
tuned in Appendix \ref{app:convert} to derive a surface temperature of
$3725\,\mbox{K}$.
Our best fit model consisted of fractional luminosities from the two
components such that,
\begin{eqnarray}
 \mbox{L}_{total} & = & 0.85\,\mbox{L}_{warm}\;+\;0.15\,\mbox{L}_{cool}.
\label{eq:app:ltot}
\end{eqnarray}
Taking $\mbox{M}_{V}\,=\,7.89$ for HII~3063, we can derive the $V$ band
contribution of each component by writing 
\begin{eqnarray}
 \mbox{M}_{V,warm} - 7.89 & = & -2.5\;\log\,(\,0.85\,\mbox{L}_{tot}\,) \\
 \mbox{M}_{V,cool} - 7.89 & = & -2.5\;\log\,(\,0.15\,\mbox{L}_{tot}\,),
\label{eq:app:mageq}
\end{eqnarray}
and normalizing such that $\mbox{L}_{total} = 1$, we find that
$\mbox{M}_{V,warm}\,=\,8.07$ and $\mbox{M}_{V,cool}\,=\,9.95$. Using the tables
from \citet{sk82} and \citet{bessell91} the $V-I_{C}$~values for Gl~3 and Gl~382 
correspond to bolometric corrections of -0.57 and -1.65, respectively, and yield
bolometric luminosities of $\mbox{M}_{bol,warm}\,=\,7.5$ and $\mbox{M}_{bol,warm}\,=\,8.3$.
With these values we can write
\begin{eqnarray}
 \mbox{M}_{bol,cool} - \mbox{M}_{bol,warm} & = & 0.80 \\
  & = & -2.5\;\log\,(\,\mbox{L}_{warm}/\mbox{L}_{cool}\,), 
\label{eq:app:magdiff}
\end{eqnarray}
such that the resulting ratio of the luminosity of the warm and cool components is
\begin{eqnarray}
 \mbox{L}_{warm}/\mbox{L}_{cool} & = & 2.09
\label{eq:app:ratio}
\end{eqnarray}

Since,
\begin{eqnarray}
 \mbox{L}_{} & \sim & 4\,\pi\;\mbox{R}^{2}\;\sigma\;\mbox{T}^{4}
\label{eq:app:lbol}
\end{eqnarray}
we can write,
\begin{eqnarray}
 \mbox{L}_{warm}/\mbox{L}_{cool} & = & (\mbox{A}_{warm}/\mbox{A}_{cool})^{2} * (\mbox{T}_{warm}/\mbox{T}_{cool})^{4},
\label{eq:app:lbol2}
\end{eqnarray}
and substituting the effective temperatures for the field dwarfs and the
luminosity ratio from Equation \ref{eq:app:ratio}, we find that the ratio
of the surface areas of the cool and warm components is:
\begin{eqnarray}
 \mbox{A}_{warm}/\mbox{A}_{cool} & = & 0.97,
\label{eq:app:answer}
\end{eqnarray}
or equal surface areas for the cool and warm components.

\section{Converting Theoretical Isochrones into $BVI$ Color-Magnitude Diagrams}
\label{app:convert}

\begin{figure} %
    \includegraphics[angle=00,totalheight=6.5in]{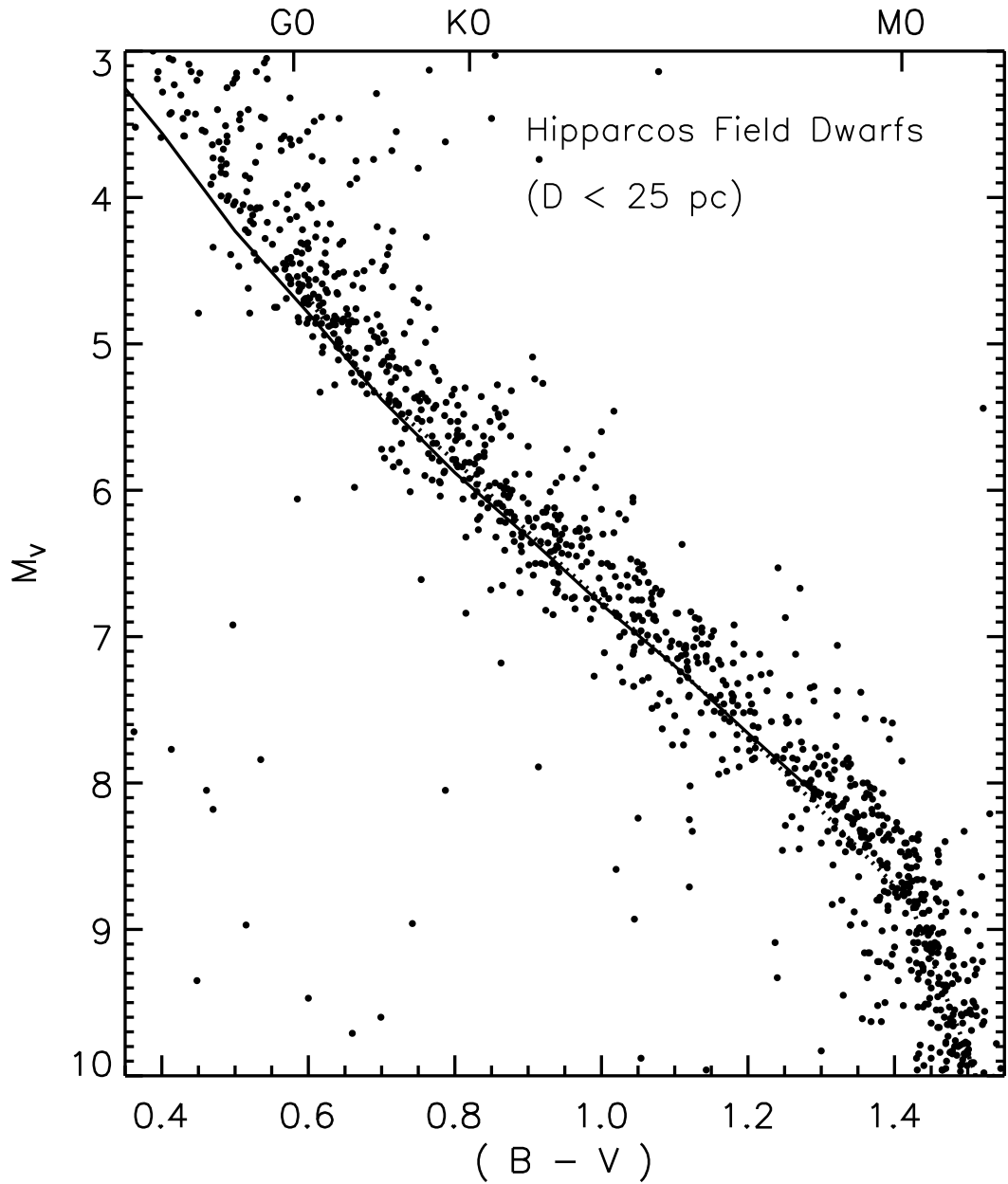}
    \caption{
	$M_V$ versus $B-V$ color-magnitude diagram for
	nearby (D$<25$pc) field stars from the Hipparcos catalog. 
	The solid curve shown is the main sequence locus utilized
	in Figure \ref{fig:jonesbvv}, illustrating the 
	apparent difference between the field K dwarfs and
	the Pleiades stars of the same absolute magnitude.
	The dotted curve is the 2 Gyr isochrone from \citet{dm97}
	converted to observables using the color-temperature relations
	defined in Appendix \ref{app:convert}; note its agreement with
	the \citeauthor{ji58} MS and with the Hipparcos field dwarfs
	through the M dwarfs.
    \label{fig:hippbvv}
    }
\end{figure}

\begin{figure} %
    \centering
    \includegraphics[angle=00,width=2.1in]{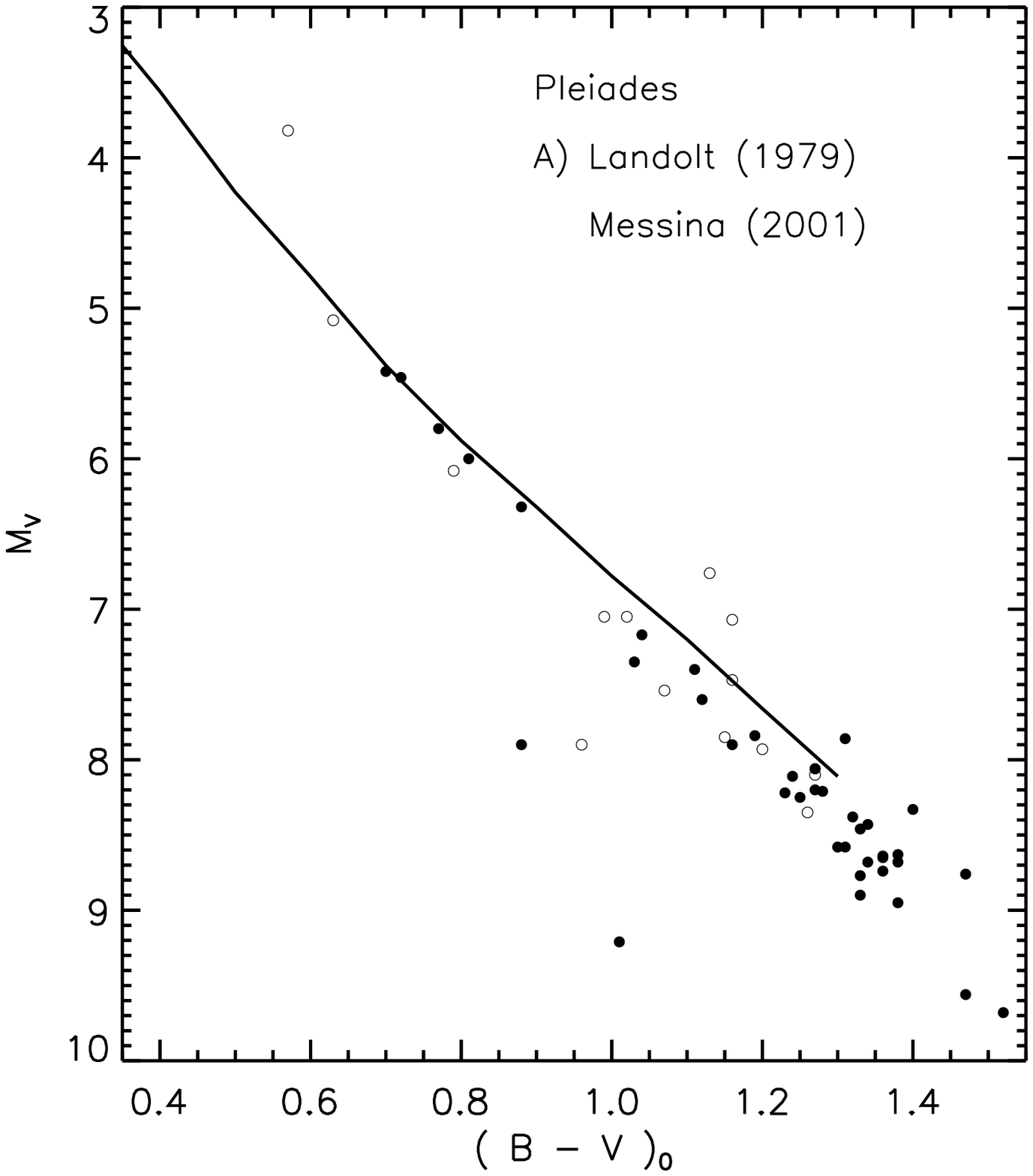}%
    \hspace{0.08in}%
    \includegraphics[angle=00,width=2.1in]{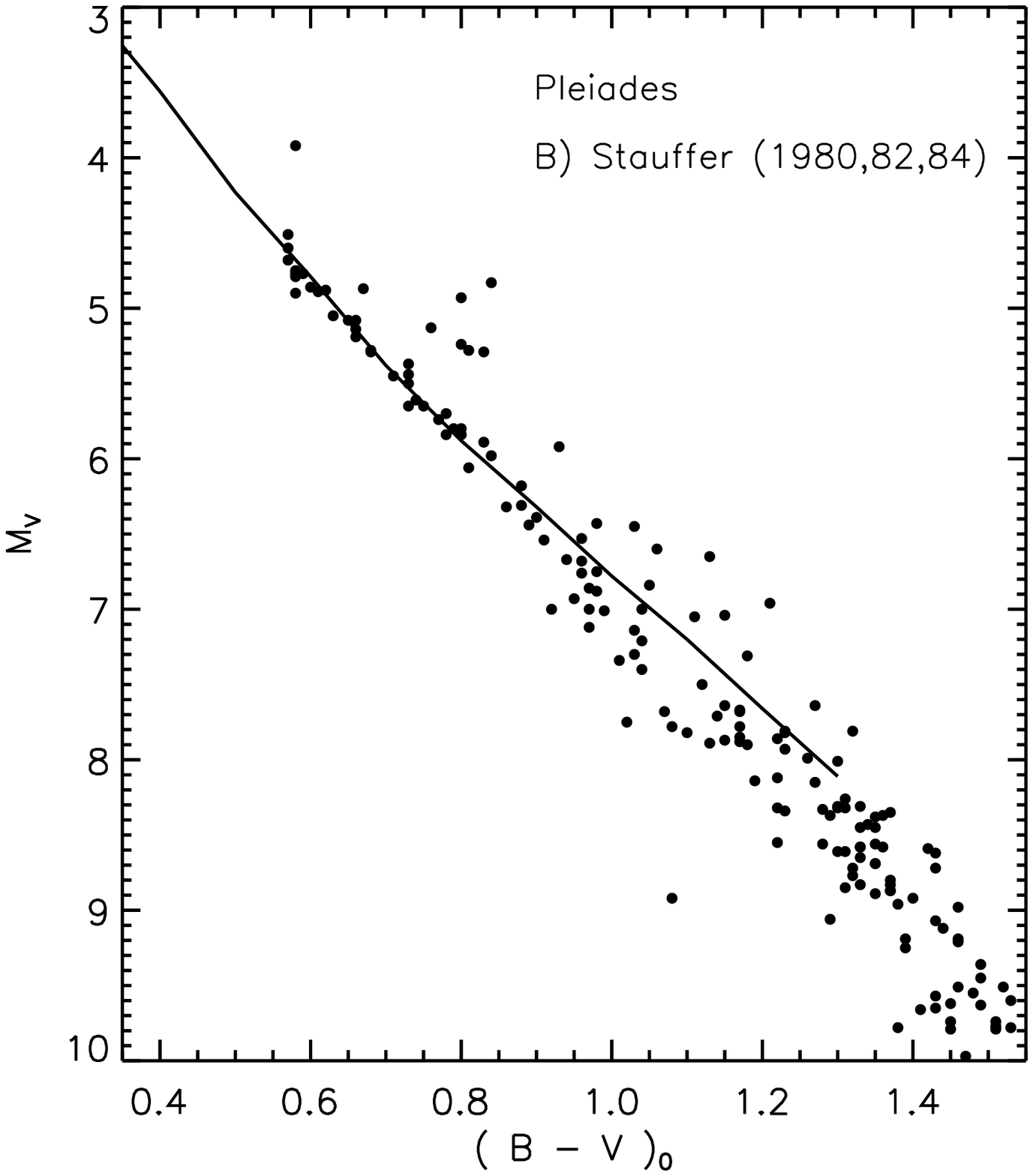}%
    \hfill 
    \includegraphics[angle=00,width=2.1in]{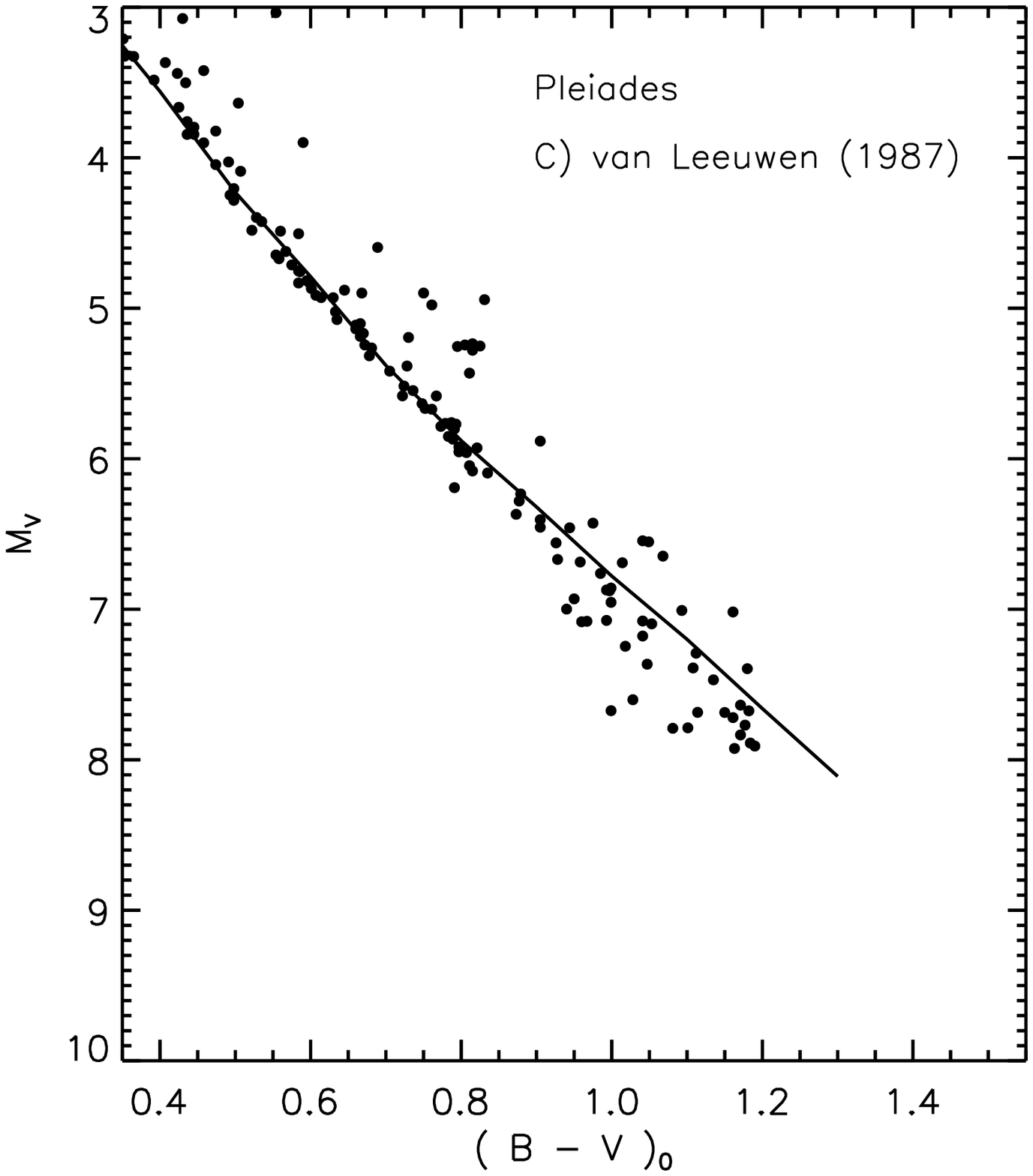}%
    \hspace{0.08in}%
    \includegraphics[angle=00,width=2.1in]{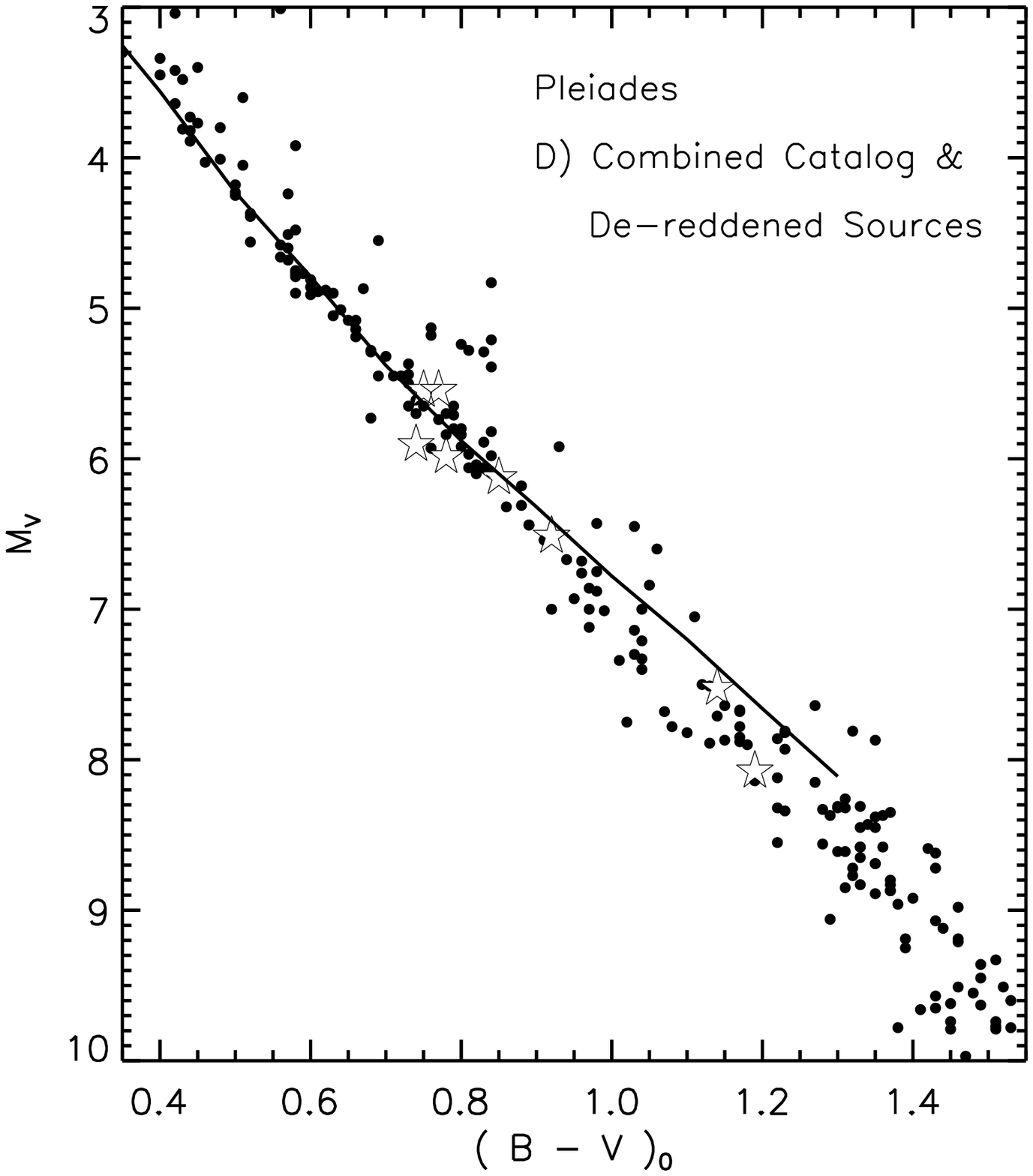}
    \hfill 
    \caption{
	Pleiades photometry from independent datasets:
        A) Landolt 1979 \& Messina 2001; B) Stauffer 1980, 1982, 1984;
        C) van Leeuwen 1987 (converted from the observed Walraven VBLUW
  	photometry into the Johnson system. See text).
	The stars from Messina (2001) are shown in (A) with open circles.
	The displacement of the K dwarfs relative to the field
	star locus is the same in all cases, making it very
	unlikely this is due to a systematic error in photometric
	calibration. D) Combined Pleiades catalog used for the reminder 
	of this paper.  De-reddened sources with large extinctions
	($E(B-V) > 0.2)$) are shown as large stars.
	Note that the majority of sources remaining above the
	main sequence are unresolved binaries. See text.
	Distance modulus and $E(B-V)$ for the Pleiades are the same
	as used in Figure \ref{fig:jonesbvv}.
    \label{fig:otherbvv}
    }
\end{figure}

\begin{figure} %
    \centering
    \includegraphics[angle=00,width=2.1in]{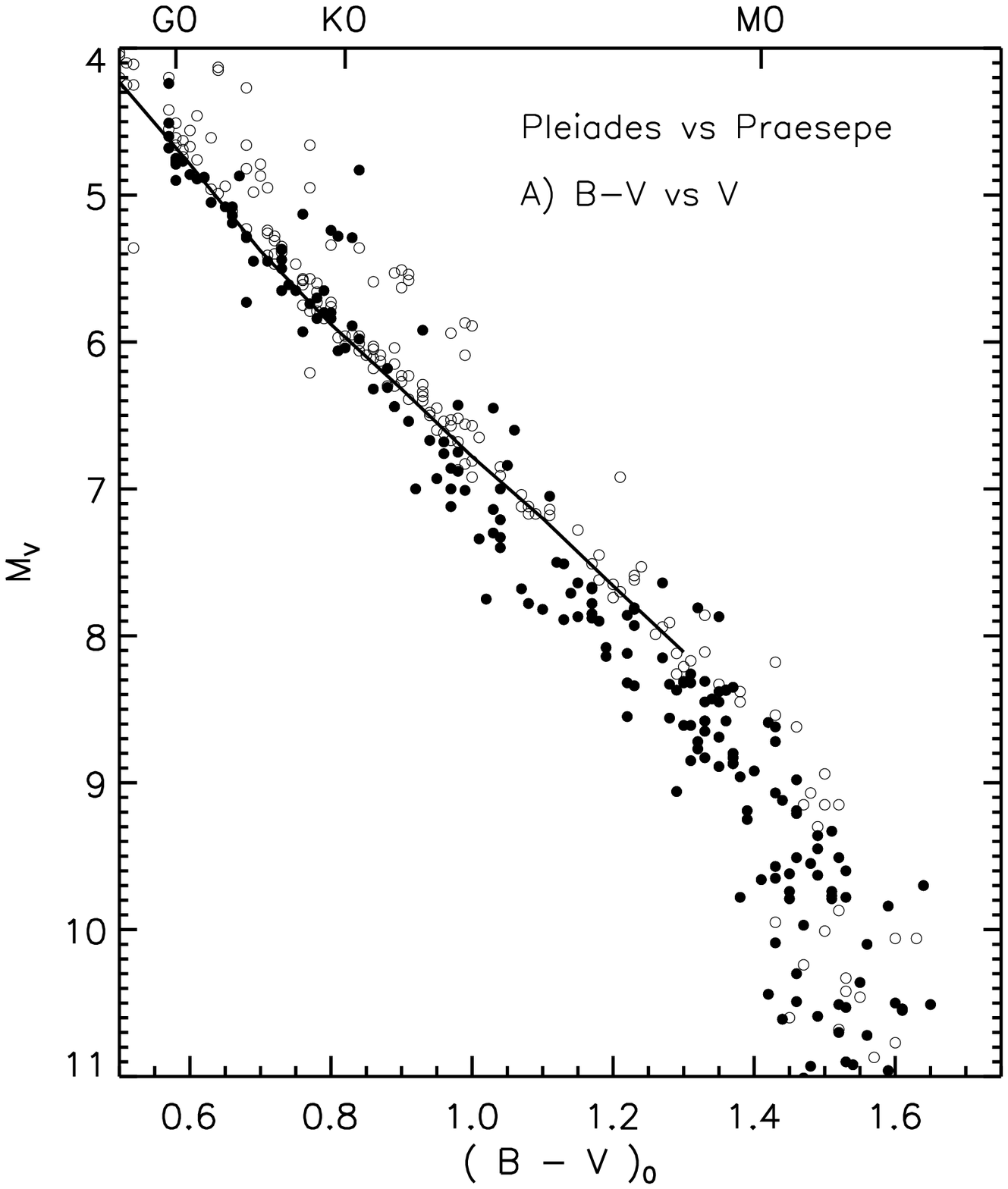}%
    \hspace{0.08in}%
    \includegraphics[angle=00,width=2.1in]{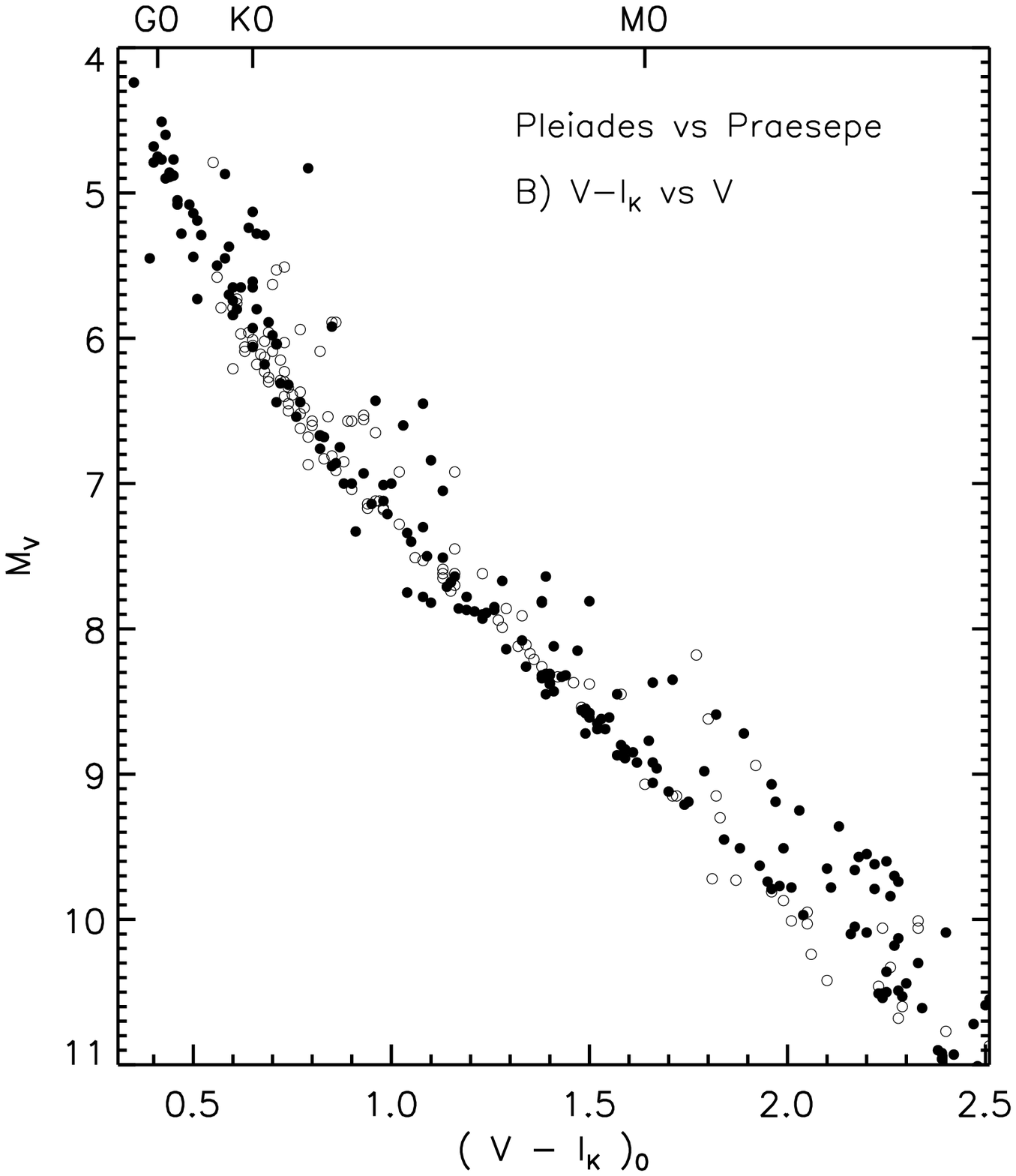}%
    \hspace{0.08in}%
    \includegraphics[angle=00,width=2.1in]{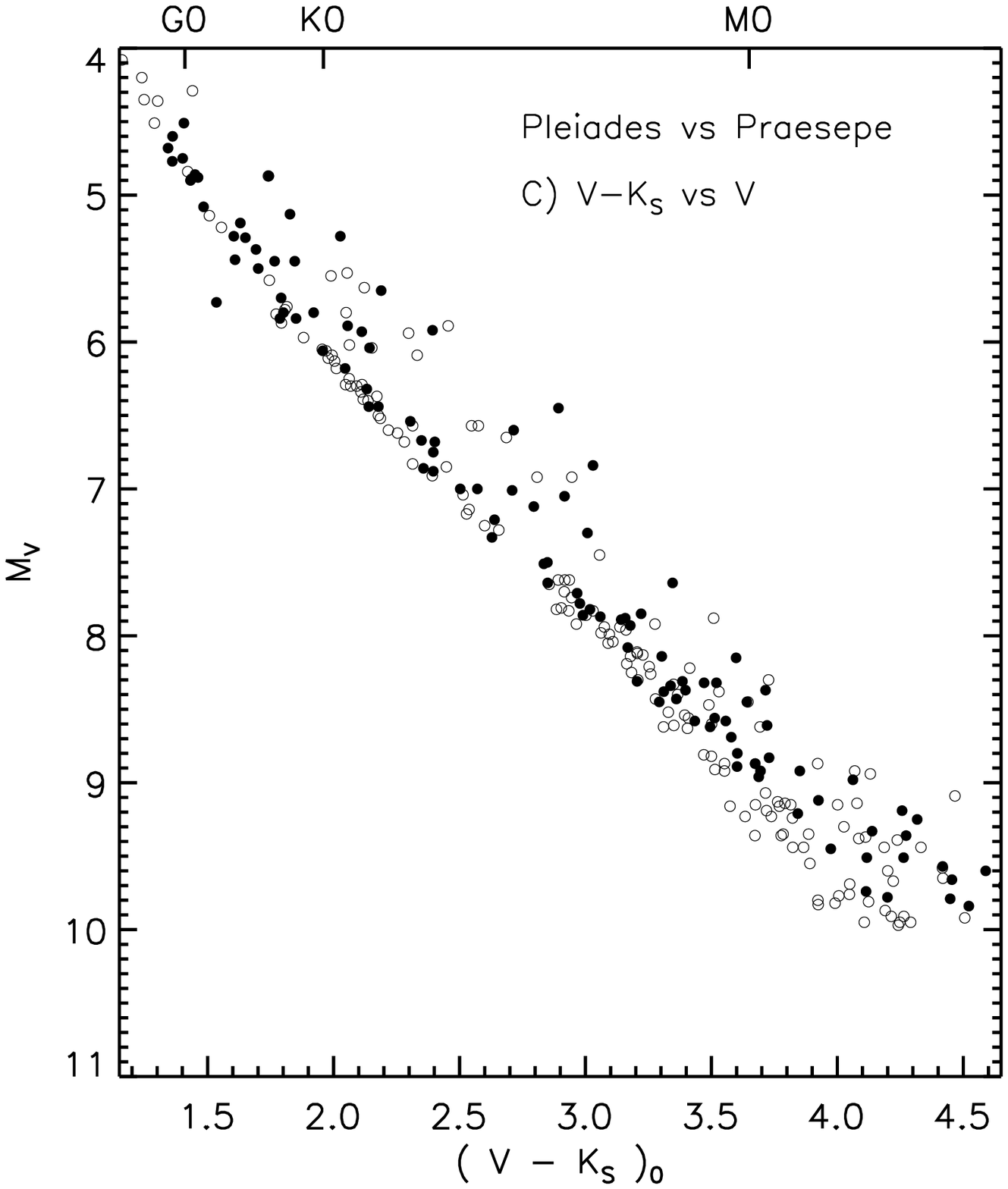}
    \caption{
	Comparison of Pleiades (filled circles) and Praesepe
	(open circles) color-magnitude (CM) diagrams.
    	A) CM diagram comparing the low mass stars in the Pleiades and
 	Praesepe using $B-V$ as the color index. Note the
   	displacemnt of the Pleiades locus below Praesepe;
	Note that fainter than $M_V\,=\,9$, increased photometric error
	in the $B-V$ color probably precludes any meaningful comparison
	between these two cluster sequences.
	B) CM diagram comparing the Pleiades and Praesepe
	low mass stars using $V-I_K$ as the color index.  There is no
	obvious systematic difference between the two clusters
	in this plane, suggesting that the anomaly illustrated in
        Figure \ref{fig:jonesbvv} and panel A is due mostly to a blue
	excess in the $B$ band.
	C) CM diagram comparing the Pleiades and Praesepe low
	mass stars using $V-K_S$ as the color index.  The fainter
	Pleiades stars now lie systematically brighter than
	the main-sequence locus defined by the Praesepe stars.
    \label{fig:plecmds}
    }
\end{figure}

\begin{figure} %
    \includegraphics[angle=00,totalheight=6.5in]{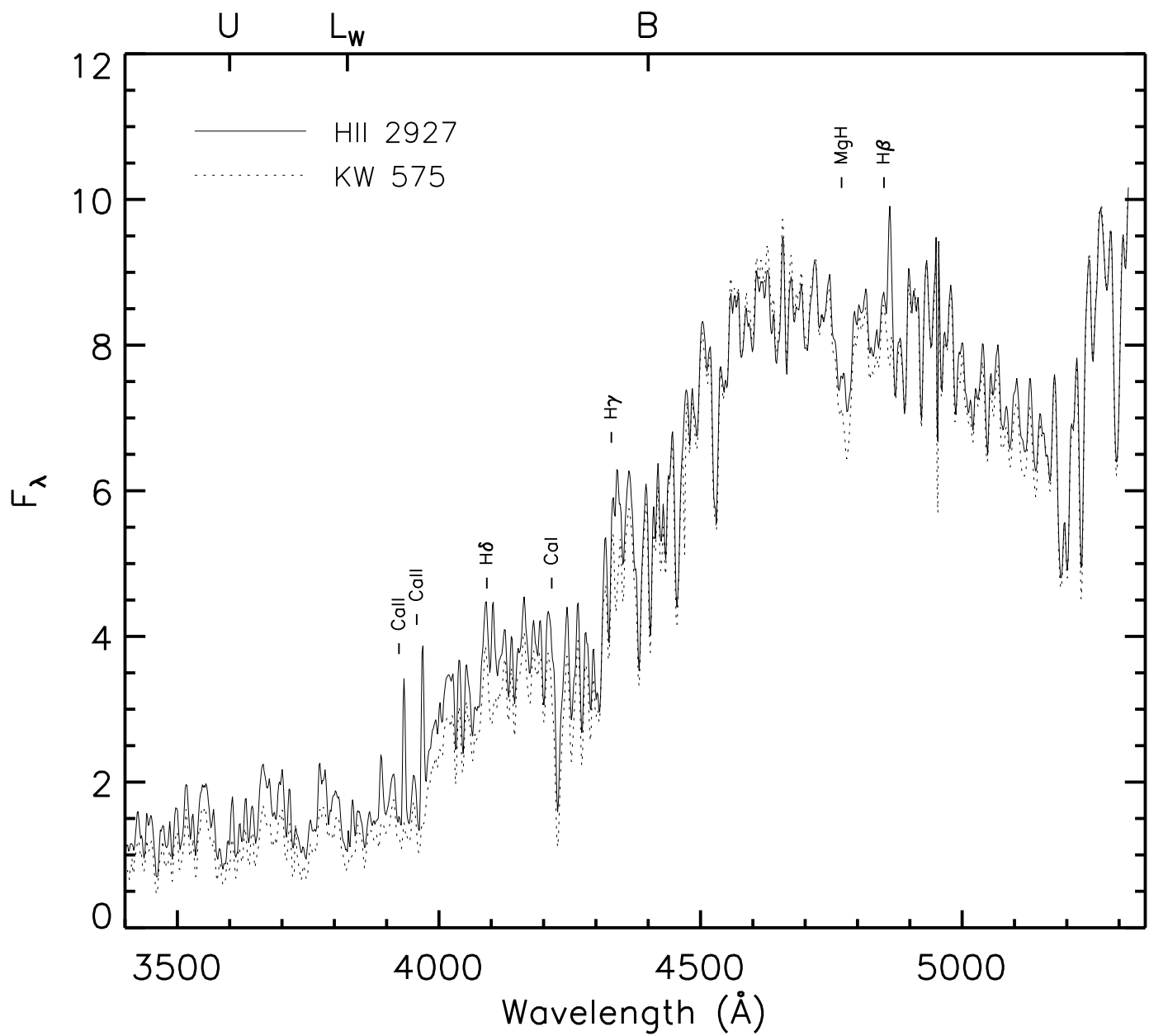}
    \caption{
	Spectra from the blue channel of LRIS for a Pleiades (HII~2927)
 	and Praesepe (KW~575) K dwarf whose $V-I_K$ colors are essentially
	identical. The spectra were normalized at 5300 \AA.
        The centers of broadband Johnson $UB$ and Walraven $L$
	filters are shown along the top axis of this figure.
    \label{fig:lris}
    }
\end{figure}

\begin{figure} %
    \centering
    \includegraphics[angle=00,width=3.0in]{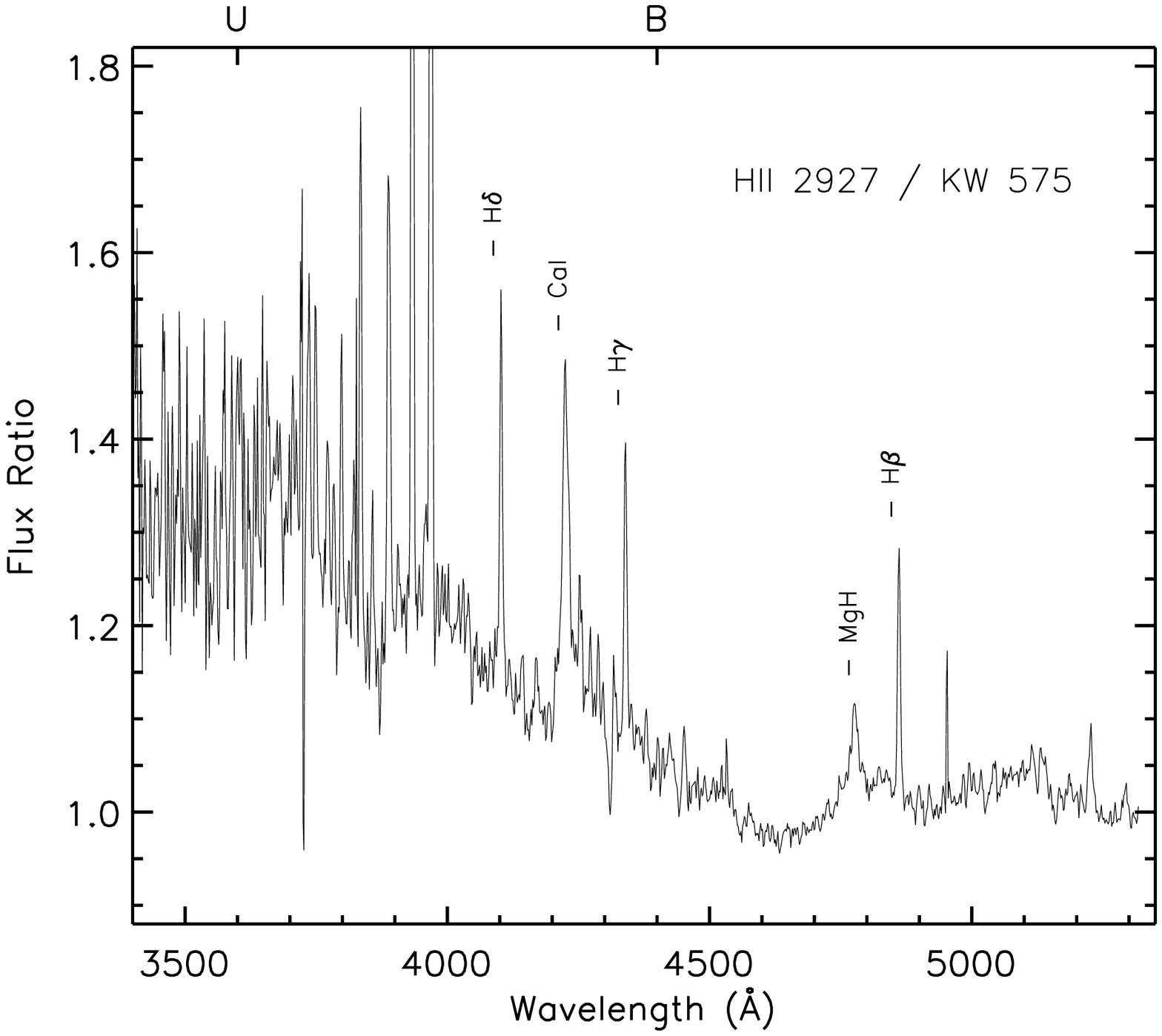}%
    \hspace{0.1in}%
    \includegraphics[angle=00,width=3.0in]{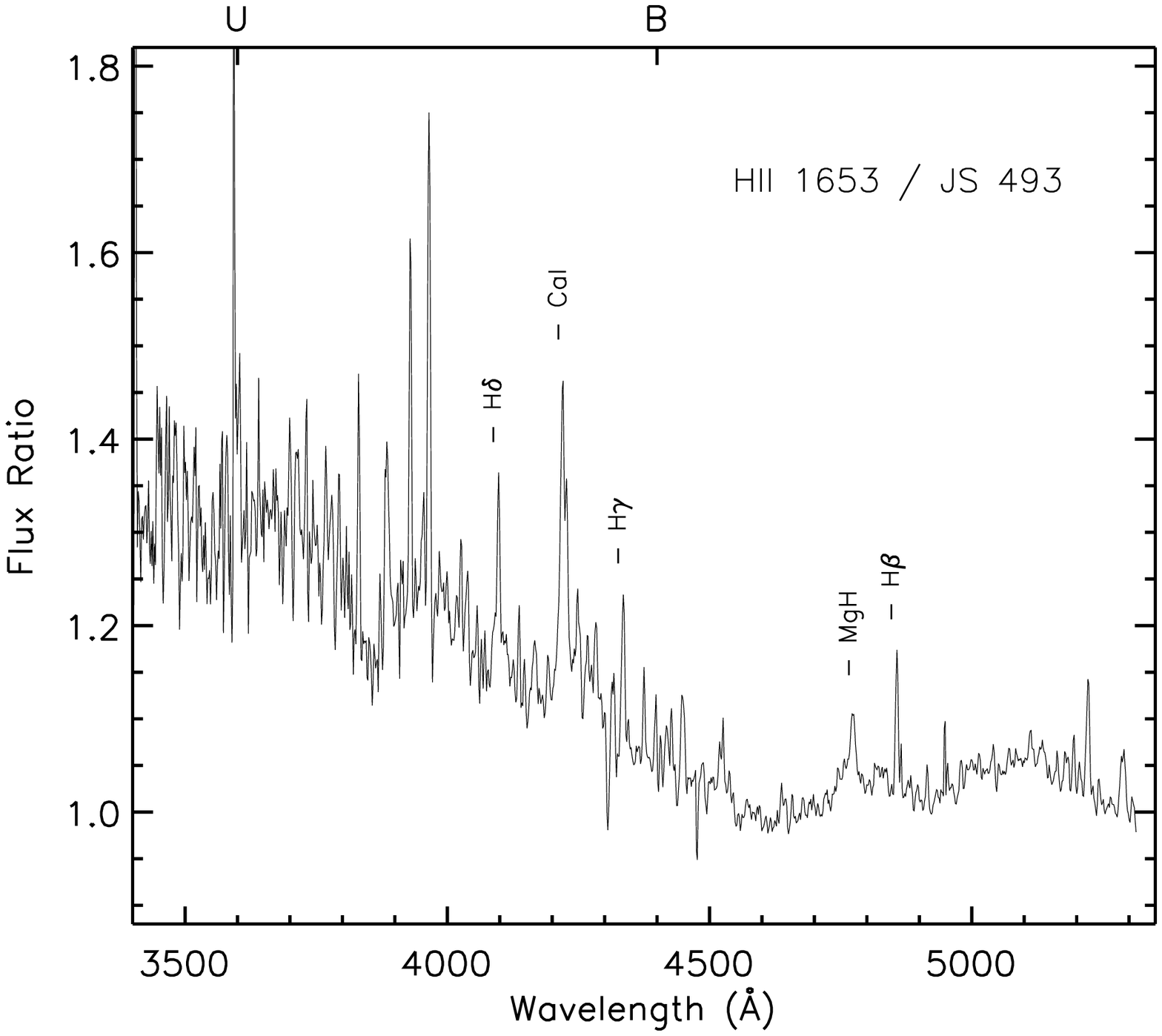}
    \caption{
	Ratio spectra showing the result of dividing
	each Pleiades K dwarf spectrum by the spectrum of its
	matched Praesepe K dwarf, after normalizing at 5300 \AA.
	See text for a description of the features present.
    \label{fig:normblue}
    }
\end{figure}

\begin{figure} %
    \centering
    \includegraphics[angle=00,width=3.0in]{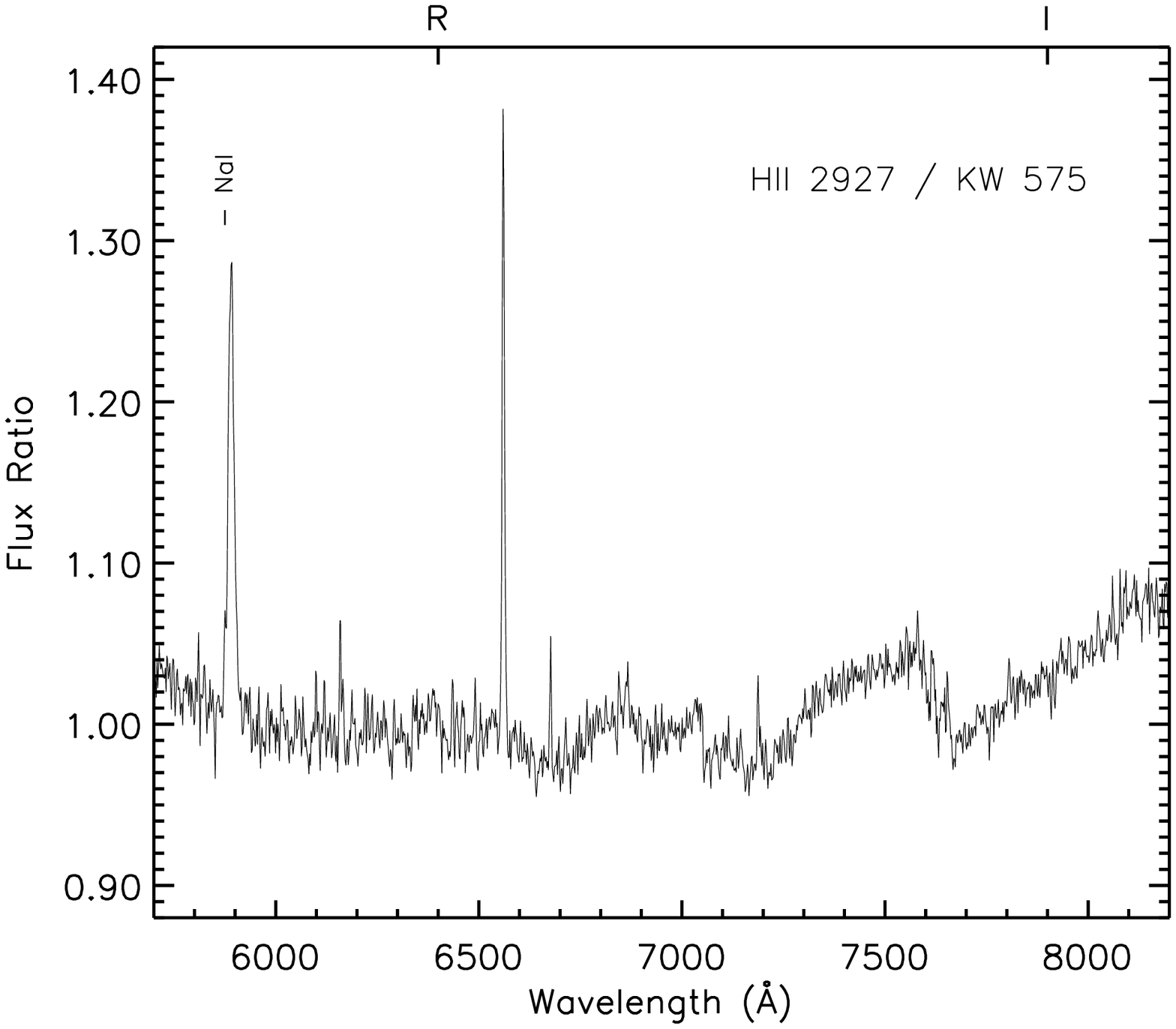}%
    \hspace{0.1in}%
    \includegraphics[angle=00,width=3.0in]{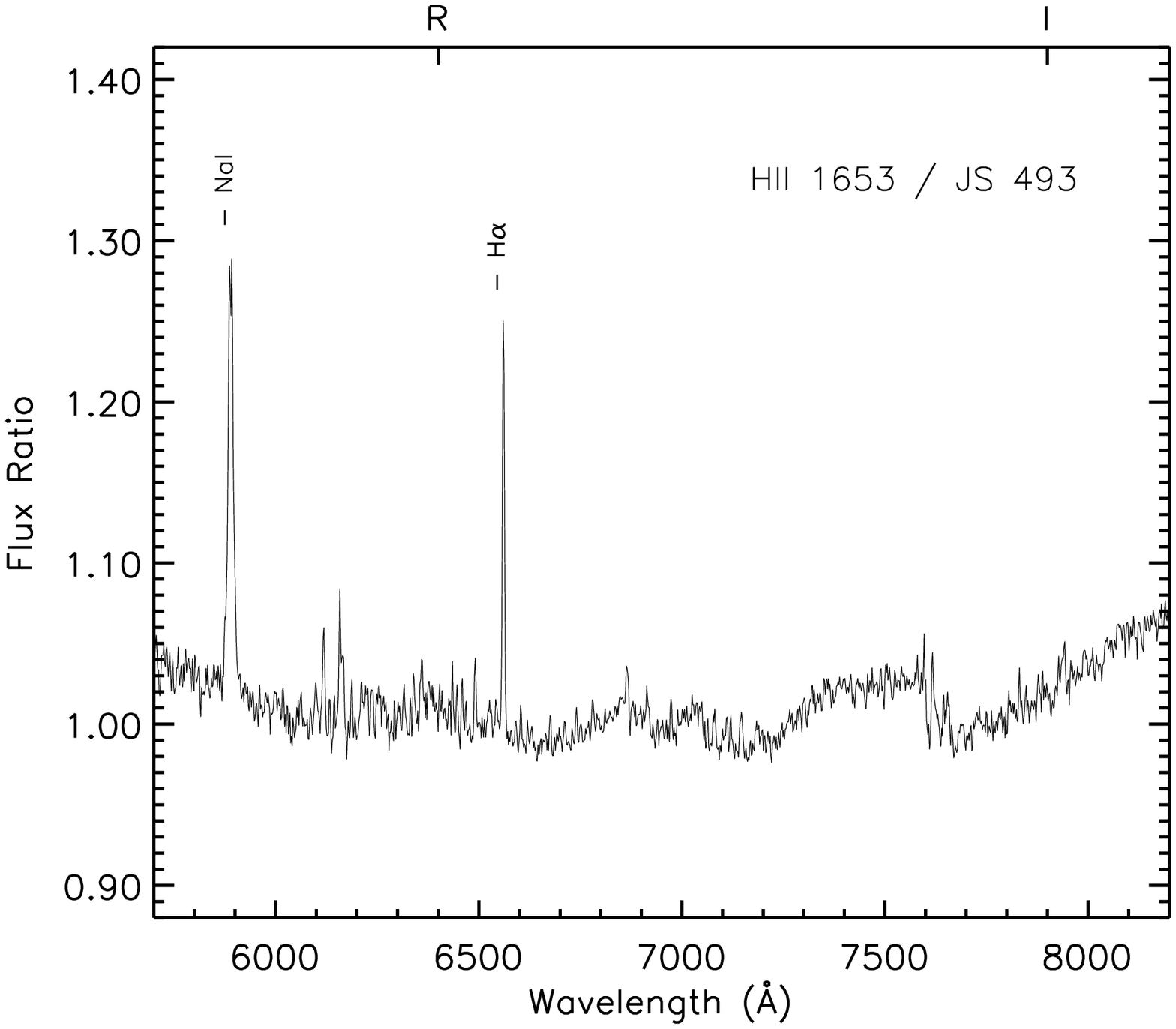}
    \caption{
	Same as for Figure \ref{fig:normblue}, except using the spectra
	obtained with the red arm of the LRIS, and where in this
	case the normalization was at 6000 \AA.
    \label{fig:normred}
    }
\end{figure}

\begin{figure} %
    \includegraphics[angle=00,totalheight=6.5in]{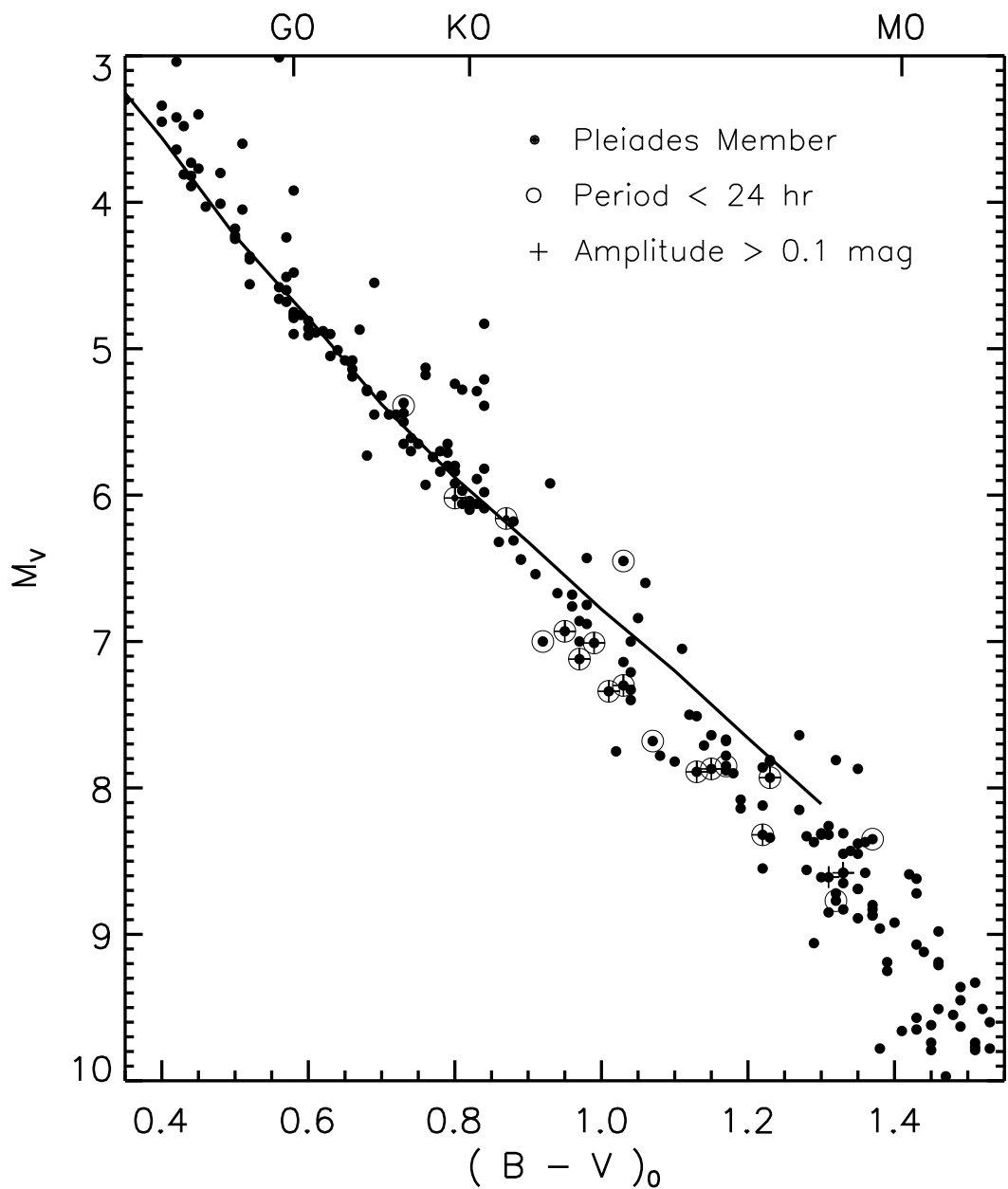}
    \caption{
	CM diagram for the Pleiades using photometry from
	Stauffer (1980,1982, 1984).  Special symbols are used to
	indicate the most rapidly rotating K dwarfs and those with
	the largest $V$ band photometric light curve amplitudes.
    \label{fig:rotvar}
    }
\end{figure}

\begin{figure} %
    \includegraphics[angle=00,width=5in]{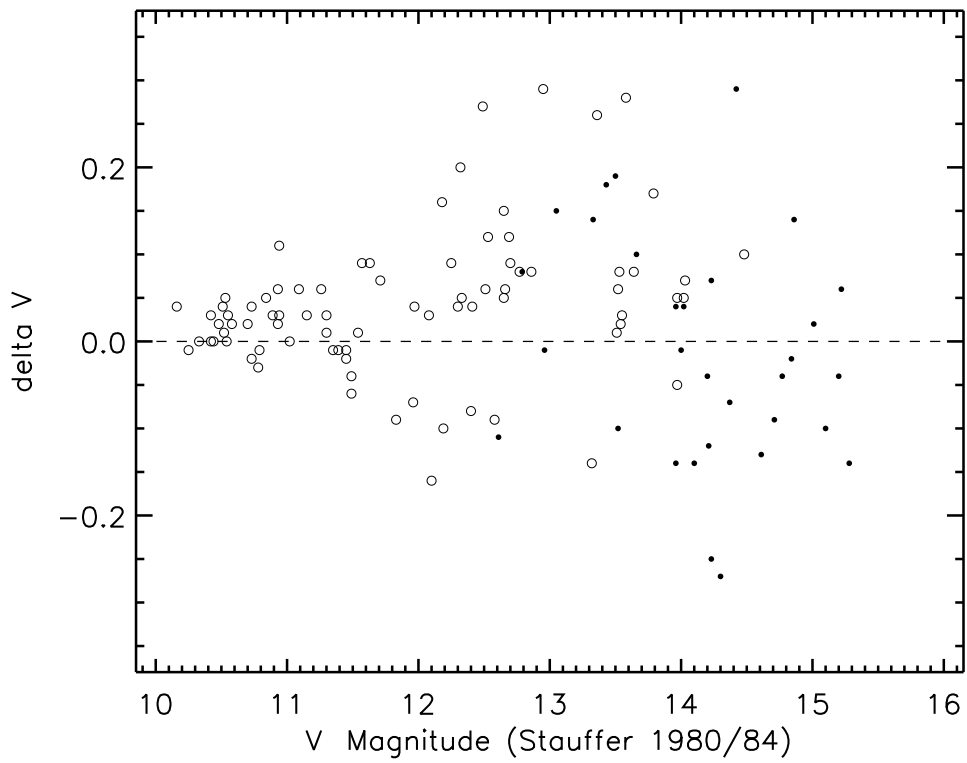}
    \caption{
	The difference in $V$ magnitude as measured in 1958 by Johnson
	\& Mitchell and as measured in the 1980's by Stauffer,
	versus the 1980's $V$ magnitude.   The fainter stars from
	Johnson \& Mitchell only had photographic photometry -
	expected to have lower intrinsic accuracy; those stars
	are shown with different symbols (solid points).
    \label{fig:deltav}
    }
\end{figure}

\begin{figure} %
    \includegraphics[angle=00,width=5in]{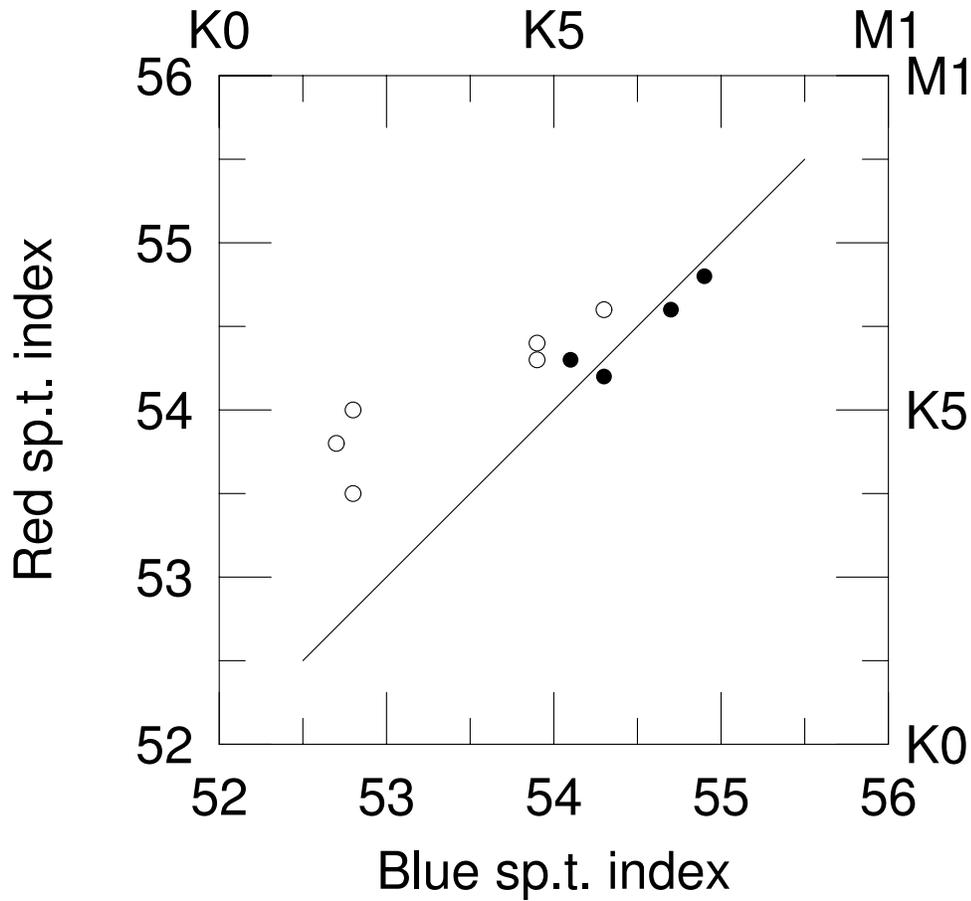}
    \caption{
	Comparison of spectral types derived by independently
	fitting the blue and red LRIS spectra of individual
	Pleiades (open circles) and Praesepe (filled circles)
	sources to the spectral library of \citet{jhc84}.
	While Praesepe sources have inferred spectral types
	independent of wavelength regime, younger Pleiades dwarfs
	display blue channel spectral types that are systematically
	earlier than those inferred from the red channel spectra.
    \label{fig:spts}
    }
\end{figure}

\begin{figure} %
    \includegraphics[angle=00,width=5in]{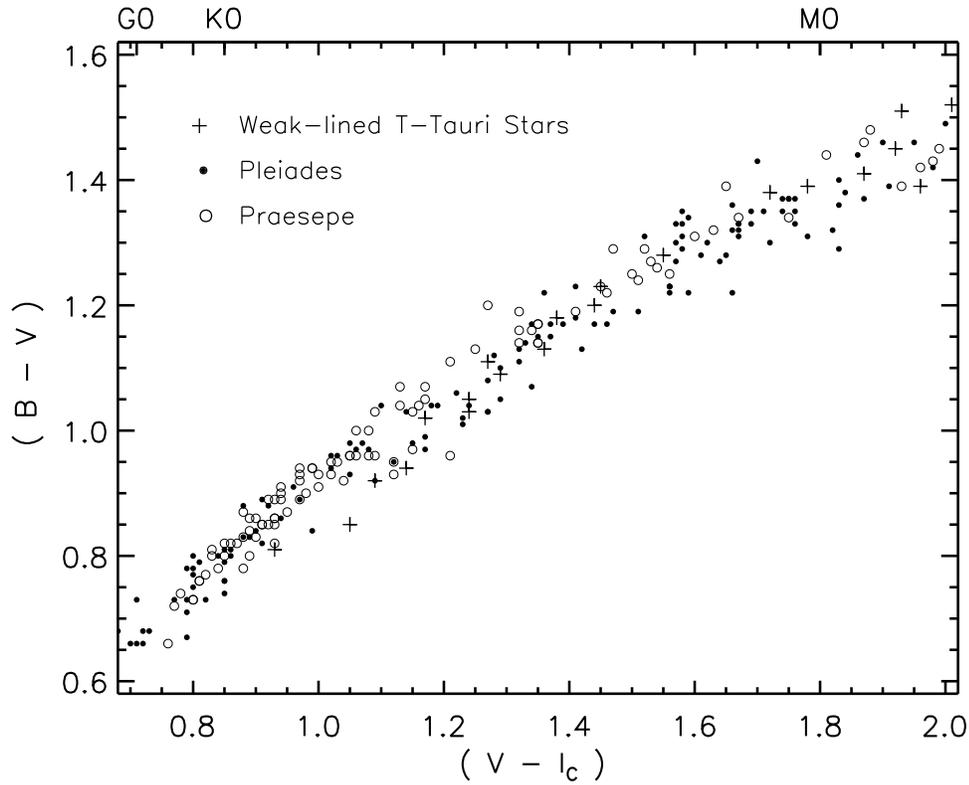}
    \caption{
	Comparison of the colors of weak-lined T-Tauri stars (WTTs)
	to low mass stars in the Pleiades and Praesepe.  
	The Pleiades K dwarf anomaly is evidenced in this figure
	by the fact that the later type Pleiades stars are systematically
	bluer in $B-V$ (for a given $V-I$) compared to the Praesepe locus.
	The fact that the WTTs fall amongst the Pleiades stars
	in this color range ($V-I > 1.1$) suggests that the
	physical mechanism causing the unusual spectral energy
	distributions for the Pleiades K dwarfs also affects the WTTs.
    \label{fig:wtts}
    }
\end{figure}

\begin{figure} %
    \centering
    \includegraphics[angle=00,width=3.0in]{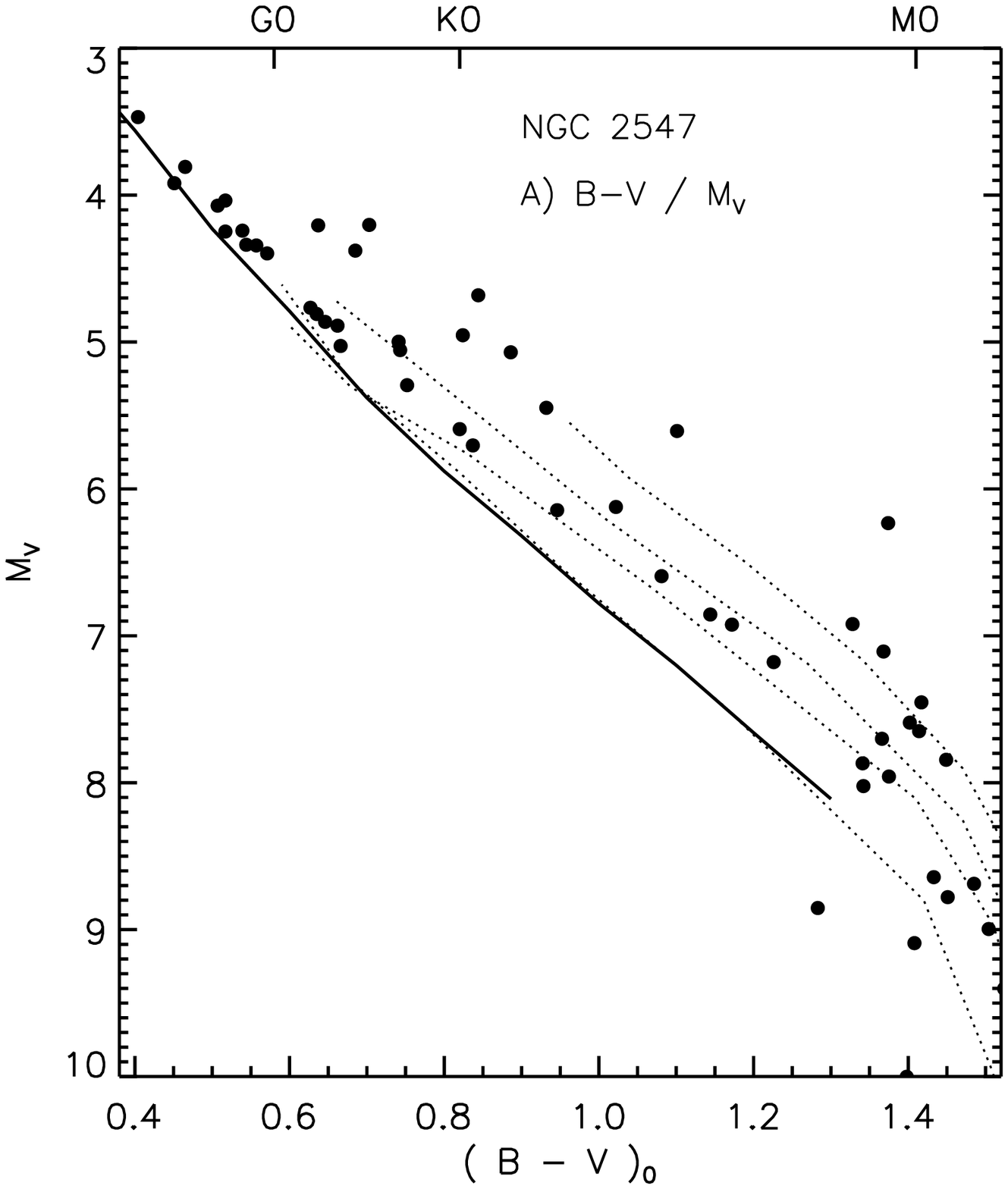}%
    \hspace{0.1in}%
    \includegraphics[angle=00,width=3.0in]{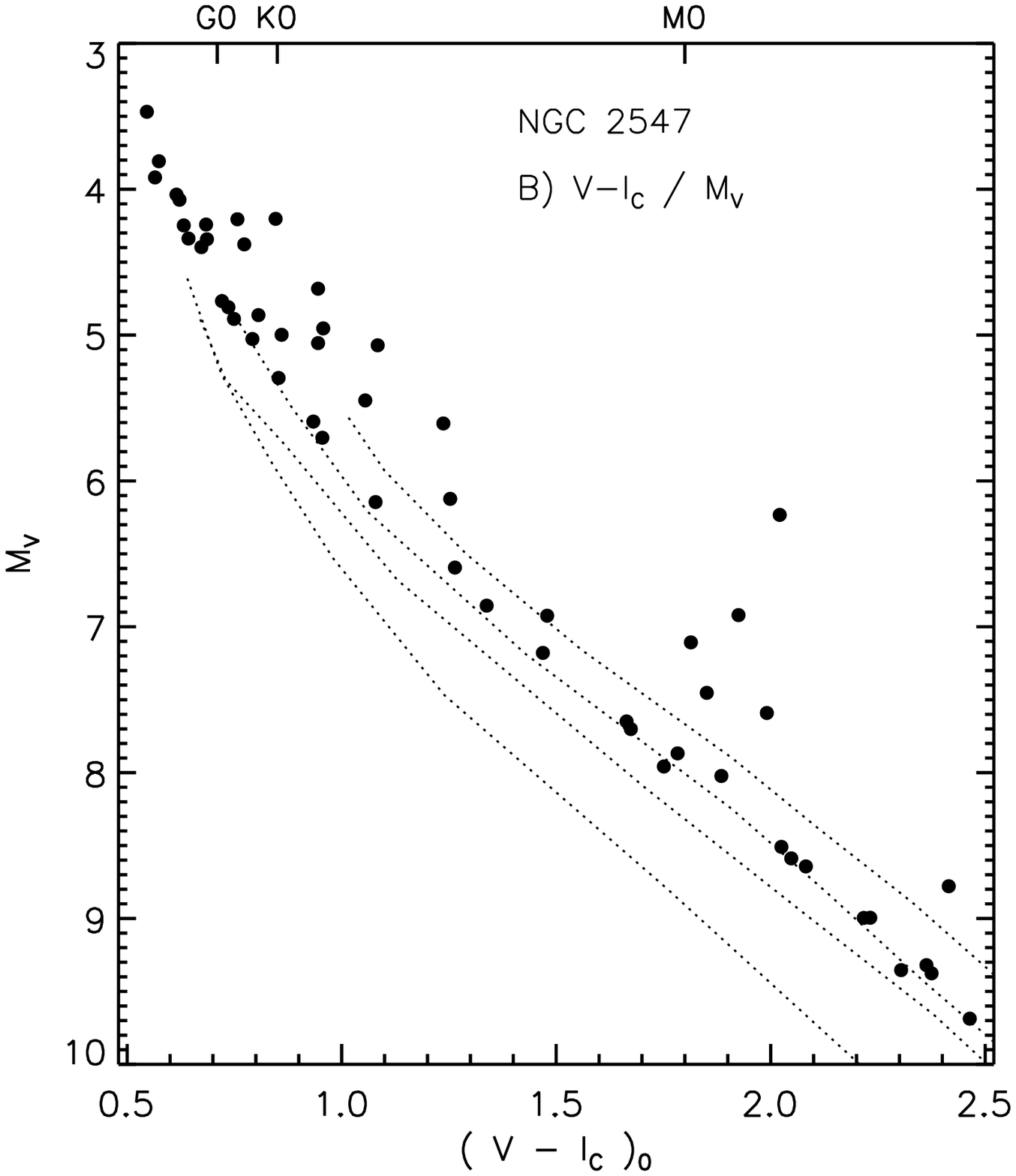}
    \caption{
	Comparison of candidate members \citep[X-ray selected,][]{jeffries98}
        of NGC~2547 to theoretical isochrones for a young ($\tau < 50$ Myr)
	pre-main sequence cluster. This figure shows that the inferred
	contraction age of NGC~2547 depends upon the temperature surragote
	used, akin to the Pleiades in Figure \ref{fig:plecmds}.
	Ages derived using $B-V$ are 5-10 Myr older than those derived using
	$V-I$ and for this cluster the effect extends from G to K type stars. 
	Shown are isochrones (dashed lines) for 20, 30 and 40 Myr and for
	2 Gyr from \citet{dm97} converted into observable quantities using
	the relations defined in Appendix \ref{app:convert}. 
	The Johnson \& Iriarte MS is also shown as a solid line in (A).
    \label{fig:n2547}
    }
\end{figure}

\begin{figure} %
    \includegraphics[angle=00,width=6.0in]{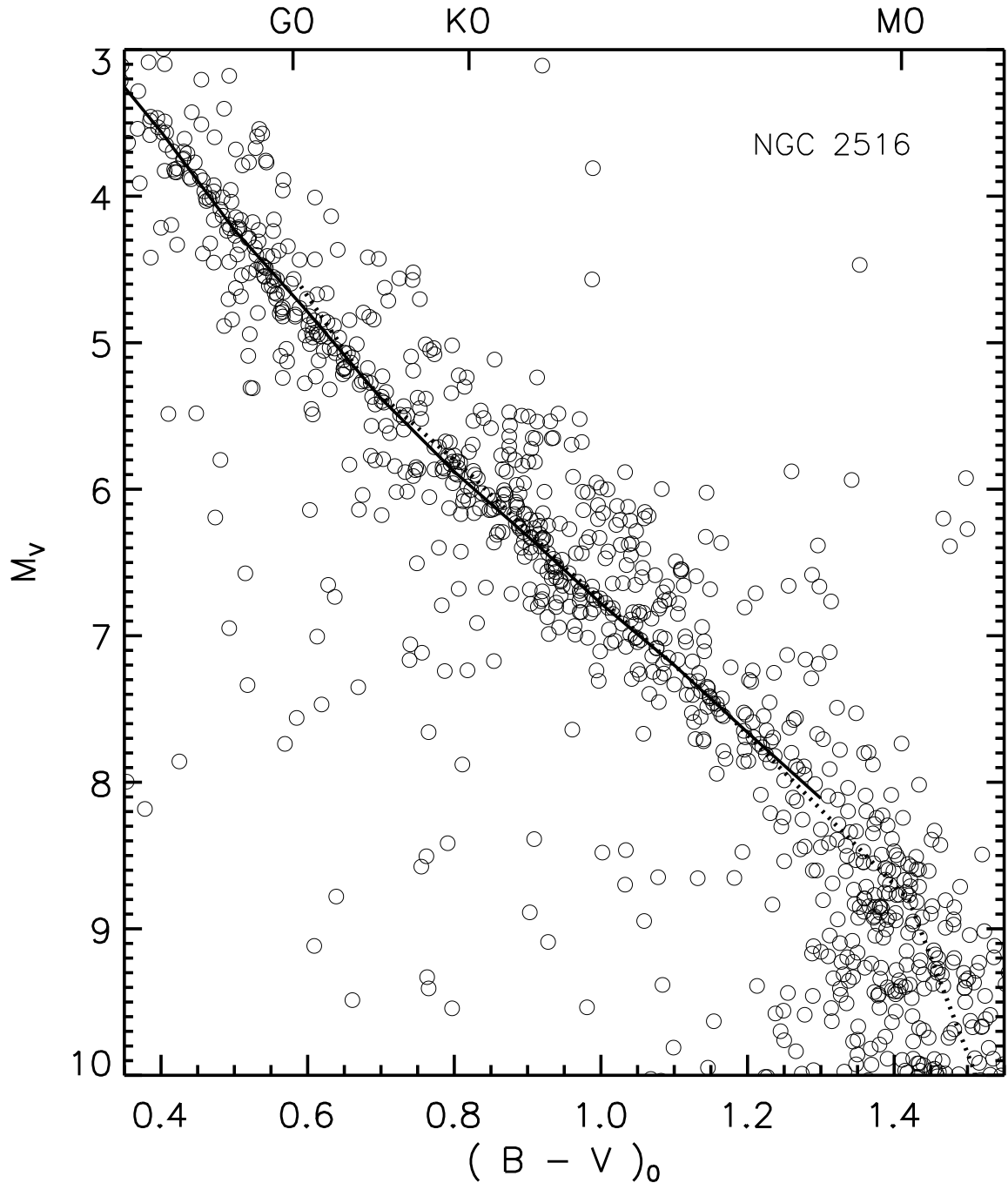}
    \caption{
    	Color-magnitude diagram for NGC~2516, as derived from the photometry
	of \citet{jeffries01} for sources filtered on their $V-I$ versus $V$
	CMD.  This $\sim$200 Myr old cluster may or may not show the Pleiades anomaly
	for K dwarfs, or the anomaly may have evolved to later type M dwarfs.
	Solid and dotted curves are the Johnson \& Iriarte MS and the 
	2 Gyr isochrone from \citet{dm97}, respectively.
    \label{fig:n2516}
    } 
\end{figure}

\begin{figure} %
    \includegraphics[angle=00,width=6.0in]{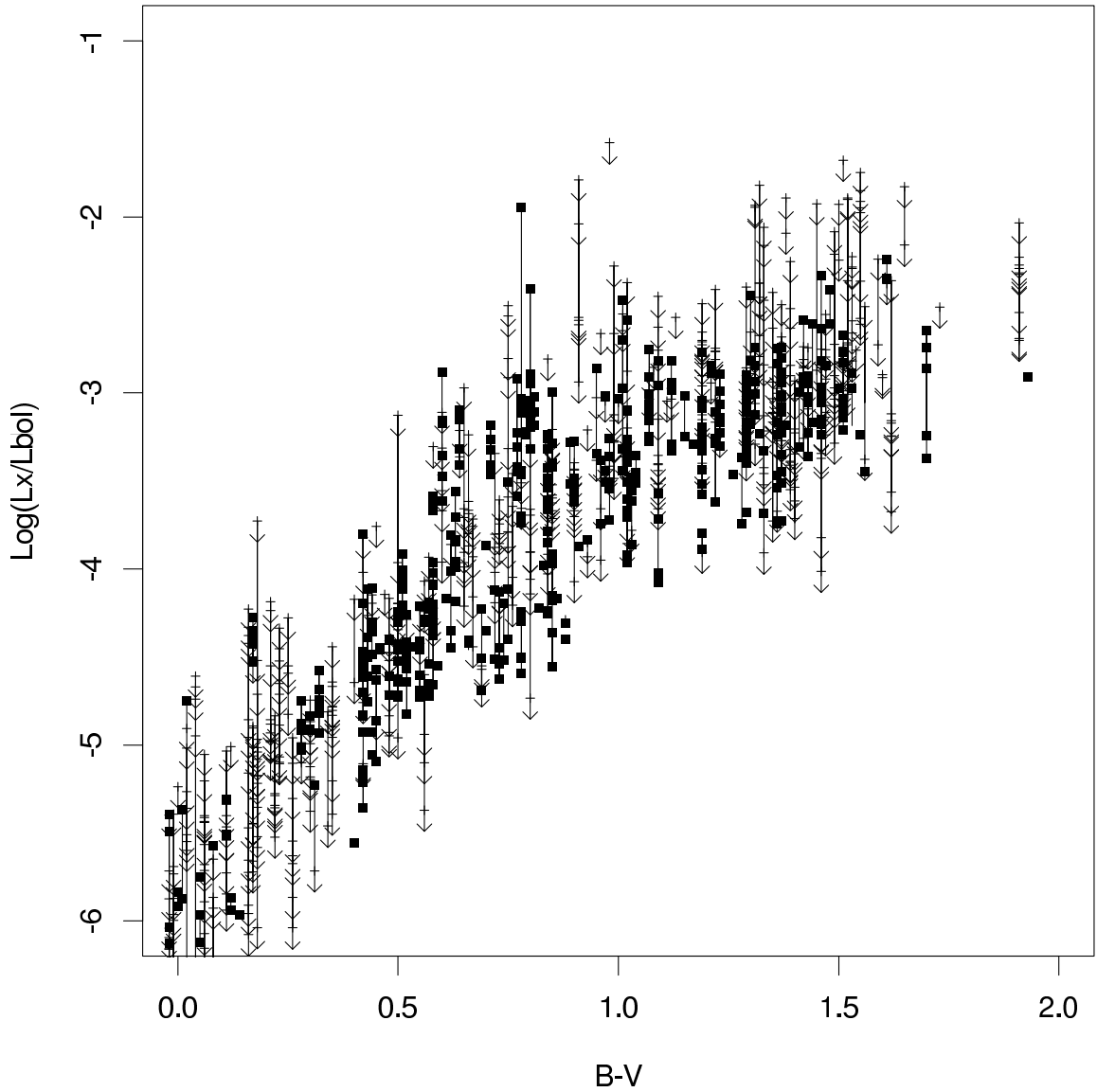}
    \caption{
	Normalized X-ray flux for Pleiades members from ROSAT
	observations. Repeat observations are connected by 
	vertical lines. Updated from \citet[][Micela, private
	communication]{micela99}. See text for a description.
    \label{fig:xray}
    }
\end{figure}

\begin{figure} %
    \includegraphics[angle=00,width=6.0in]{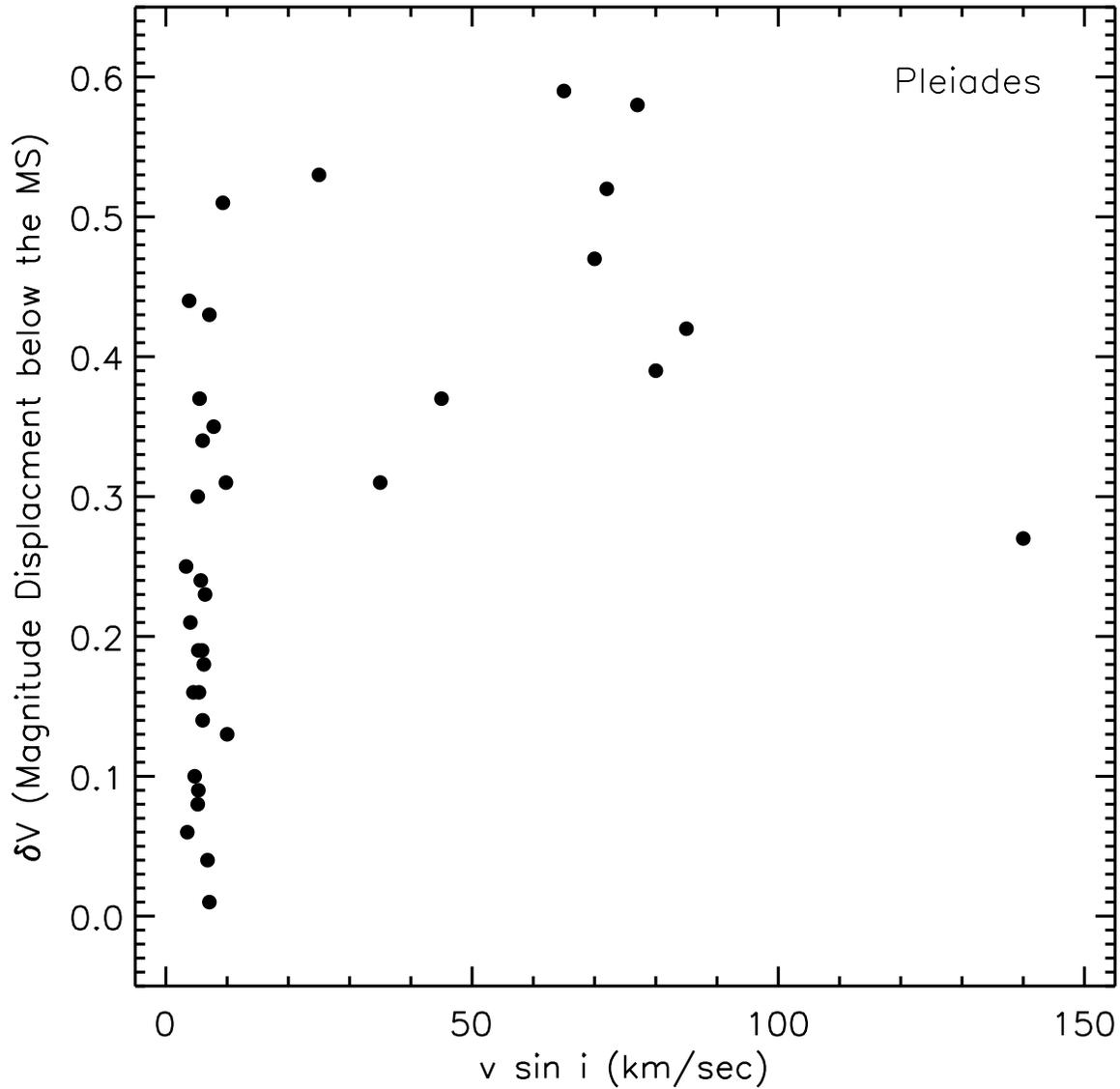}
    \caption{
	Dependence of the Pleiades K dwarf anomaly on ${v}\,\sin i $.	
	The displacement of the individual Pleiades members away
	from the Johnson \& Iriarte MS is measured in delta $V$ magnitude.
	The plot includes stars with $0.9\,<\,B-V\,<\,1.2$.
    \label{fig:vsini}
    }
\end{figure}

To convert the fundamental stellar parameters ($\mbox{L}_{\sun}$,
$\mbox{T}_{eff}$) provided by the \citet{dm97} theoretical models into
observable quantities, we employed prescription(s) modified from those 
used in \citet{shb95} and \citet{stauffer97}. In those works the theoretical
isochrones were converted using relationships between effective temperature
and color tuned to match certain observational constraints.
In this work we tuned these conversions to match two general constraints:
the Pleiades lower main sequence in $V-I\,/\,V$ CMD and the Hipparcos
field stars in the $B-V\,/\,V$ CMD (see Figure \ref{fig:hippbvv}).

First, we tuned the 100 Myr isochrone of \citet{dm97} to
fit the lower main sequence of the Pleiades, extracting a
$V-I\mbox{ to T}_{eff}$ relationship from this fit. 
This is an update from previous works where we used the
70 Myr isochrone.
This relationship was then used to convert the theoretical tracks
into the $V-I\,/\,V$ CMD for ages ranging from 10 Myr to 2 Gyr.

Next, we used the Gliese star database of \citet{bessell90}
to derive a correspondence between $V-I$ and $B-V$ for 
present day (rather than zero-age) field dwarfs.
To derive a zero-age conversion from $V-I$ to $B-V$
we supposed that the ``young'' 2 Gyr isochrone of \citeauthor{dm97}
should lie near the lower envelope to the field star locus in
Figure \ref{fig:hippbvv}. Thus, we tuned the $V-I$ to $B-V$
correspondence we derived from the Gliese stars
until the 2 Gyr $B-V\,/\,V$ isochrone fit both 
the \citeauthor{ji58} MS and provided a lower boundary to the
Hipparcos delineated nearby field dwarfs. This latter fit
constraint is shown in Figure \ref{fig:hippbvv}.



\clearpage 

\begin{deluxetable}{lccccc} 
\tablewidth{0pt}
\tablecaption{K Dwarfs Observed with LRIS and
    Keck \label{tab:lris} } 

\tablehead{
 \colhead{Name} & 
 \colhead{$M_{V,0}$} & 
 \colhead{$(B - V)_0$} & 
 \colhead{$(V - I_{K})_{0}$} &
 \colhead{$(V - I_{C})_{0}$} &
 \colhead{Cluster}
} 

\startdata
HII  324   &  7.34 & 1.01 & 1.04 & 1.23 & Pleiades  \\
HII  882   &  7.30 & 1.03 & 1.08 & 1.27 & Pleiades  \\ 
HII 1653   &  7.85 & 1.17 & 1.26 & 1.44 & Pleiades  \\
HII 1883   &  7.01 & 0.99 & 0.99 & 1.17 & Pleiades  \\
HII 2927   &  8.32 & 1.22 & 1.38 & 1.56 & Pleiades  \\ 
HII 3063   &  7.89 & 1.13 & 1.24 & 1.42 & Pleiades  \\
                                   \\
JS 267     &  7.70 & 1.21 & 1.16 & 1.35 & Praesepe  \\
JS 473     &  7.65 & 1.20 & 1.13 & 1.32 & Praesepe  \\
JS 493     &  7.94 & 1.27 & 1.27 & 1.45 & Praesepe  \\ 
KW 575     &  8.26 & 1.29 & 1.38 & 1.56 & Praesepe  \\ 
\enddata 

\end{deluxetable}

\clearpage 

\begin{deluxetable}{lcccccc}
\tablewidth{0pt}
\tablecaption{Pleiades K Dwarfs and Model Spotted 
    Stars \label{tab:models}}

\tablehead{
 \colhead{Name}  &
 \colhead{$M_B$} & 
 \colhead{$M_V$} &
 \colhead{$M_{I(C)}$} & 
 \colhead{$M_J$} & 
 \colhead{$M_H$} &
 \colhead{$M_K$}
 }

\startdata 
HII 2927                &  9.54 &  8.32 &  6.76 &  5.66 &  5.02 &  4.84  \\ 
model\tablenotemark{(1)}&  9.54 &  8.32 &  6.76 &  5.65 &  5.04 &  4.84  \\ 
                        &       &       &       &       &       &        \\
HII 3063                &  9.02 &  7.89 &  6.47 &  5.52 &  4.87 &  4.74  \\ 
model\tablenotemark{(2)}&  9.01 &  7.89 &  6.47 &  5.53 &  4.90 &  4.72  \\ 
                        &       &       &       &       &       &        \\
HII 2908                &  8.92 &  7.82 &  6.53 &  5.45 &  4.94 &  4.79  \\
model\tablenotemark{(3)}&  8.93 &  7.82 &  6.52 &  5.47 &  4.95 &  4.78  \\ 
                        &       &       &       &       &       &        \\
HII 1305                &  9.02 &  7.87 &  6.50 &  5.58 &  4.95 &  4.80  \\
model\tablenotemark{(4)}&  9.02 &  7.87 &  6.48 &  5.58 &  4.96 &  4.79  \\ 
                        &       &       &       &       &       &        \\
\enddata
\tablenotetext{(1)}{model fit -- $25\%$ GL 149;~~ $75\%$ GL 382;~IR shift=0.12}
\tablenotetext{(2)}{model fit -- $15\%$ GL 3;~~~~ $85\%$ GL 382;~IR shift=0.01}
\tablenotetext{(3)}{model fit -- $15\%$ GL 293.2; $85\%$ GL 7;~~~IR shift=0.15}
\tablenotetext{(4)}{model fit -- $15\%$ GL 3;~~~~ $85\%$ GL 334;~IR shift=0.00}
\end{deluxetable}

\clearpage 

\begin{deluxetable}{llcccc} 
\tablewidth{0pt}
\tablecaption{Metallicity Determinations for the Pleiades and
the Praesepe \label{tab:app:metals} } 

\tablehead{
 \multicolumn{2}{c}{Pleiades} &
 \multicolumn{2}{c}{Praesepe} &
 \colhead{Method}   & 
 \colhead{Ref.}     \\ 
 \colhead{$[Fe/H]$} & 
 \colhead{$\pm$}    & 
 \colhead{$[Fe/H]$} &
 \colhead{$\pm$}    &
 \colhead{}         & 
 \colhead{}         
} 

\startdata
 \nodata & \nodata & $+0.038$& $0.039$ & Spectra & 1 \\
 $-0.034$& $0.024$ & \nodata & \nodata & Spectra & 2 \\
 $+0.022$& $0.062$ & $+0.092$& $0.067$ & Spectra & 3 \\
 \nodata & \nodata & $+0.132$& $0.067$ & Spectra & 4 \\ 
 $+0.003$& $0.051$ & \nodata & \nodata & Spectra & 5 \\ \hline
 $+0.01$ & $0.02$  & \nodata & \nodata & 4 color + H-$\beta$ & 6\\ 
 \nodata & \nodata & +0.14   & \nodata & 4 color + H-$\beta$ & 7\\
\enddata 

\tablerefs{
           1: \citet{fb92};\ \
           2: \citet{boesgaard90};\ \
	   3: \citet{boesgaard89};\ \
           4: \citet{boesgaard88b};\ \
           5: \citet{boesgaard88a};\ \
           6: \citet{stello01};\ \
           7: \citet{reglero91} }

\end{deluxetable}


\clearpage
\newpage

\pagestyle{myheadings}
\pagenumbering{arabic}
\setcounter{page}{1}


\clearpage
\newpage

\clearpage
\newpage

\clearpage
\newpage

\clearpage
\newpage

\clearpage
\newpage

\clearpage
\newpage

\clearpage
\newpage

\clearpage
\newpage

\clearpage
\newpage

\clearpage
\newpage

\clearpage
\newpage

\clearpage
\newpage

\clearpage
\newpage

\clearpage
\newpage

\clearpage
\newpage

\end{document}